\documentclass[12pt]{article}
\pdfoutput=1 
\usepackage{hyperref} 

\usepackage{graphicx}
\usepackage{graphics}
\usepackage{amssymb}
\usepackage{amsmath}
\usepackage{epsf}

\usepackage[parfill]{parskip}
\usepackage[matrix,arrow]{xy}
\usepackage[usenames]{color}
\input{epsf}

\newcommand{\ba}{\begin{eqnarray*}}
\newcommand{\ea}{\end{eqnarray*}}
\newcommand{\ban}{\begin{eqnarray}}
\newcommand{\ean}{\end{eqnarray}}
\newcommand{\beq}{\begin{equation*}}
\newcommand{\eeq}{\end{equation*}}
\newcommand{\beqn}{\begin{equation}}
\newcommand{\eeqn}{\end{equation}}

\newcommand{\Tr}{{\rm Tr\,}}
\newcommand{\tr}{{\rm tr\,}}

\newcommand{\IZ}{\mathbb{Z}}
\newcommand{\IC}{\mathbb{C}}
\newcommand{\IP}{\mathbb{P}}

\newcommand{\cS}{{\cal S}}

\newcommand{\cX}{{\cal X}}

\newcommand{\cO}{{\cal O}}

\newcommand{\cF}{{\cal F}}

\newcommand{\td}{\tilde}

\newcommand{\CYX}{{\mathfrak X}}

\newcommand{\Res}{\mathop{{\rm Res}\,}}
\newcommand{\Li}{\rm Li}
\newcommand{\Disc}{\mathop{{\rm Disc}\,}}

\newcommand{\spcurve}{{{\mathcal S}}}
\newcommand{\curve}{{{\mathcal C}}}
\newcommand{\spcurveMM}{{{\mathcal S}_{\rm MM}}}

\makeatletter
\@addtoreset{equation}{section}
\makeatother

\newcommand\encadremath[1]{\vbox{\hrule\hbox{\vrule\kern8pt
\vbox{\kern8pt \hbox{$\displaystyle #1$}\kern8pt}
\kern8pt\vrule}\hrule}}
\def\enca#1{\vbox{\hrule\hbox{
\vrule\kern8pt\vbox{\kern8pt \hbox{$\displaystyle #1$}
\kern8pt} \kern8pt\vrule}\hrule}}

\def \nn{\nonumber}

\newdimen\tableauside\tableauside=1.0ex
\newdimen\tableaurule\tableaurule=0.4pt
\newdimen\tableaustep
\def\phantomhrule#1{\hbox{\vbox to0pt{\hrule height\tableaurule width#1\vss}}}
\def\phantomvrule#1{\vbox{\hbox to0pt{\vrule width\tableaurule height#1\hss}}}
\def\sqr{\vbox{%
  \phantomhrule\tableaustep
  \hbox{\phantomvrule\tableaustep\kern\tableaustep\phantomvrule\tableaustep}%
  \hbox{\vbox{\phantomhrule\tableauside}\kern-\tableaurule}}}
\def\squares#1{\hbox{\count0=#1\noindent\loop\sqr
  \advance\count0 by-1 \ifnum\count0>0\repeat}}
\def\tableau#1{\vcenter{\offinterlineskip
  \tableaustep=\tableauside\advance\tableaustep by-\tableaurule
  \kern\normallineskip\hbox
    {\kern\normallineskip\vbox
      {\gettableau#1 0 }%
     \kern\normallineskip\kern\tableaurule}%
  \kern\normallineskip\kern\tableaurule}}
\def\gettableau#1 {\ifnum#1=0\let\next=\null\else
  \squares{#1}\let\next=\gettableau\fi\next}

\begin{document}

\begin{titlepage}
\begin{flushright}
IPHT
\\
LPTENS 10/23
\end{flushright}
\begin{center}
\vskip 2cm {\Huge A matrix model for the topological string II
\\ \vskip 0.2cm}
{\LARGE The spectral curve and mirror geometry}\\
\vskip 1cm {B. Eynard${}^1$, A. Kashani-Poor${}^{2,3}$, O. Marchal${}^{1,4}$}

\vskip.6cm 
{\it ${}^1$ Institut de Physique Th\'eorique,\\
CEA, IPhT, F-91191 Gif-sur-Yvette, France,\\
CNRS, URA 2306, F-91191 Gif-sur-Yvette, France.\\ \vskip0.3cm}

{\it 
$^2$ Institut des Hautes \'Etudes Scientifiques\\
Le Bois-Marie, 35, route de Chartres, 91440 Bures-sur-Yvette, France\\ \vskip0.3cm }

{\it $^3$ Laboratoire de Physique Th\'eorique de l'\'Ecole Normale Sup\'erieure, \\
24 rue Lhomond, 75231 Paris, France \\ \vskip0.3cm}

{\it ${}^4$ Centre de recherches math\'ematiques,
Universit\'e de Montr\'eal \\
C.P. 6128, Succ. centre-ville
Montr\'eal, Qu\'e, H3C 3J7, Canada.\\}

\end{center} 
\vskip 1.5cm
\begin{abstract}
In a previous paper, we presented a matrix model reproducing the topological string partition function on an arbitrary given toric Calabi-Yau manifold. Here, we study the spectral curve of our matrix model and thus derive, upon imposing certain minimality assumptions on the spectral curve, the large volume limit of the BKMP ``remodeling the B-model'' conjecture, the claim that Gromov-Witten invariants of any toric Calabi-Yau 3-fold coincide with the spectral invariants of its mirror curve. 
\end{abstract}

\end{titlepage}
\newpage

\tableofcontents

\section{Introduction}

In a previous paper \cite{part_1}, we presented a matrix model that computes the topological string partition function at large radius on an arbitrary toric Calabi-Yau manifold $\CYX$. The goal of this paper is to determine the corresponding spectral curve $\spcurve$. 

That the partition function of a matrix model can be recovered to all genus from its spectral curve was first demonstrated in \cite{eynloop1mat}. \cite{EOFg} pushed this formalism further, showing that symplectic invariants $F_g(\spcurve)$ can be defined for any analytic affine curve $\spcurve$, with no reference to an underlying matrix model. These invariants coincide with the partition function of a matrix model when $\spcurve$ is chosen as the associated spectral curve. The symplectic invariants $F_g$ satisfy many properties reminiscent of the topological string partition function \cite{eynhaeq, Orantin,CMMV,eynMgnkappa}, motivating Bouchard, Klemm, Mari\~no, and Pasquetti (BKMP) \cite{BKMP}, building on work of Mari\~no \cite{marino2}, to conjecture that $F_g(\spcurve)$ in fact coincides with the topological string partition function on the toric Calabi-Yau manifold with mirror curve $\spcurve$. BKMP successfully checked their claim for various examples, at least to low genus. The conjecture was subsequently proved in numerous special cases \cite{eynLP, MarshakovNekrasov, KlemmSulkowski, Sulkowski,Zhou, Lin}.

Bouchard and Mari\~no \cite{BouchardMarino} noticed that an infinite framing limit of the BKMP conjecture for the framed vertex, $\CYX=\mathbb C^3$, implies a conjecture for the computation of Hurwitz numbers, namely that the Hurwitz numbers of genus $g$ are the symplectic invariants of genus $g$ for the Lambert spectral curve $e^x=y\, e^{-y}$. This conjecture was proved recently by a generalization of \cite{eynLP} using a matrix model for summing over partitions \cite{BEMS}, and also by a direct combinatorial method \cite{EMS}. Matrix models and the BKMP conjecture related to toric Calabi-Yau geometries arising from the triangulation of a strip were recently studied in \cite{OSY}.

In this paper, we derive the large radius limit of the BKMP conjecture for {\it arbitrary} toric Calabi-Yau manifolds, but with one caveat: to determine the spectral curve of our matrix model, we must make several minimality assumptions along the way. To elevate our results to a rigorous proof of the BKMP conjecture, one needs to establish a uniqueness result underlying our prescription for finding  the spectral curve to justify these minimal choices. Such a uniqueness result does not exist to date.

Recall that in \cite{part_1}, we first compute the topological string partition function on a toric Calabi-Yau geometry $\CYX_0$ which we refer to as fiducial. We then present a matrix model which reproduces this partition function. Flops and limits in the K\"ahler cone relate $\CYX_0$ to an arbitrary toric Calabi-Yau 3-fold. As we can follow the action of these operations on the partition function, we thus arrive at a matrix model for the topological string on any toric Calabi-Yau 3-folds. Here, we follow the analogous strategy, by first computing the spectral curve of the matrix model associated to $\CYX_0$, and then studying the action of flops and limits on this curve.

The plan of the paper is as follows. In section \ref{fiducial_section}, we introduce the fiducial geometry $\CYX_0$ and its mirror. The matrix model reproducing the partition function on $\CYX_0$, as derived in \cite{part_1}, is a chain of matrices matrix model. It is summarized in section \ref{s_our_matrix_model} and appendix \ref{our_matrix_model}. We review general aspects of this class of matrix models and their solutions in section \ref{generalities_on_mm}. In section \ref{the_spectral_curve}, we determine a spectral curve which satisfies all specifications outlined in section \ref{generalities_on_mm}, and demonstrate that it coincides, up to symplectic transformations, with the B-model mirror of the fiducial geometry.  While in our experience with simpler models, the conditions of section \ref{generalities_on_mm} on the spectral curve specify it uniquely, we lack a proof of this uniqueness property. We thus provide additional consistency arguments for our proposal for the spectral curve in section \ref{secproofsp}. Flops and limits in the K\"ahler cone relate the fiducial to an arbitrary toric Calabi-Yau manifold. Following the action of these operations on both sides of the conjecture in section \ref{finishing_proof} completes the argument yielding the BKMP conjecture for arbitrary toric Calabi-Yau manifolds in the large radius limit.  We conclude by discussing possible avenues along this work can be extended.

\section{The fiducial geometry and its mirror} \label{fiducial_section}

\subsection{The fiducial geometry}
\begin{figure}[h]
\centering
\includegraphics[width=10cm]{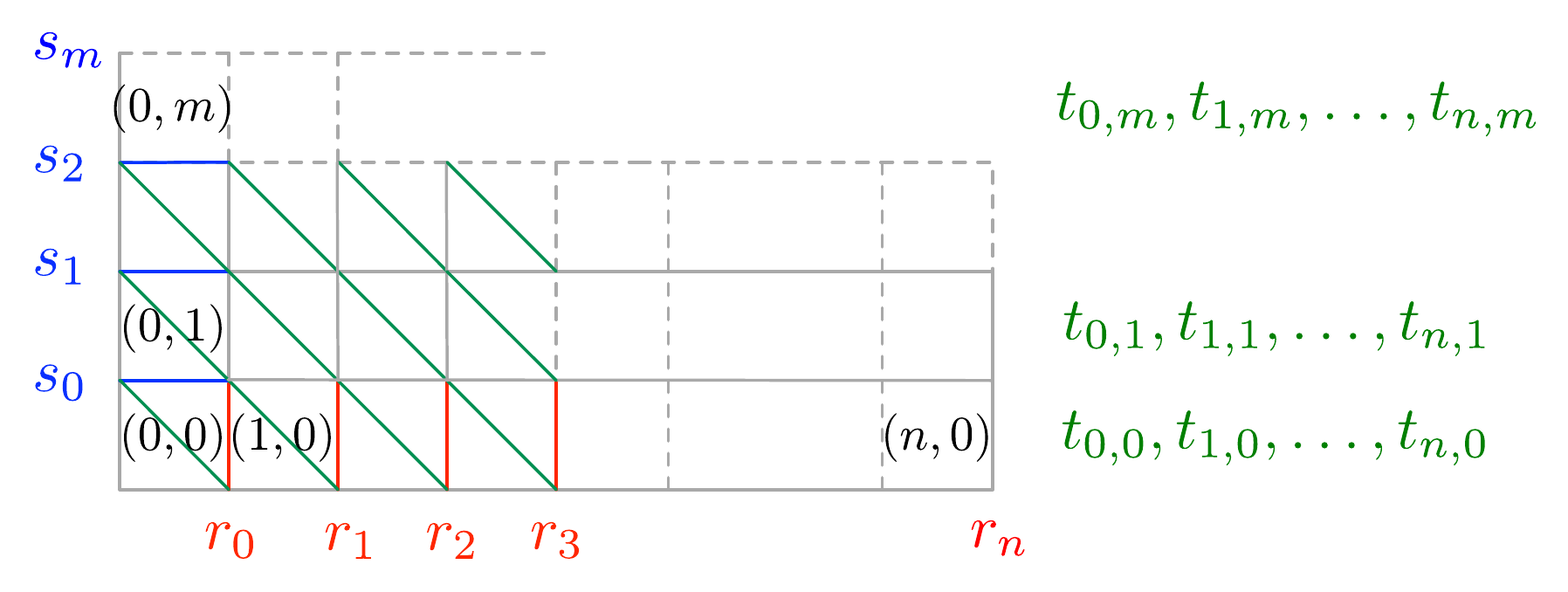}
\caption{\footnotesize{Fiducial geometry $\CYX_0$ with boxes numbered and choice of basis of $H_2(\CYX_0, \IZ)$.}}
\label{fiducial_geometry}
\end{figure}

In \cite{part_1}, we derived a matrix model reproducing the topological string partition function on the toric Calabi-Yau geometry $\CYX_0$ whose toric fan is depicted in figure \ref{fiducial_geometry}. We refer to $\CYX_0$ as our fiducial geometry; we will obtain the partition function on an arbitrary  toric  Calabi-Yau manifolds by considering flops and limits of $\CYX_0$.

\begin{figure}[h]
\centering
\includegraphics[width=3cm]{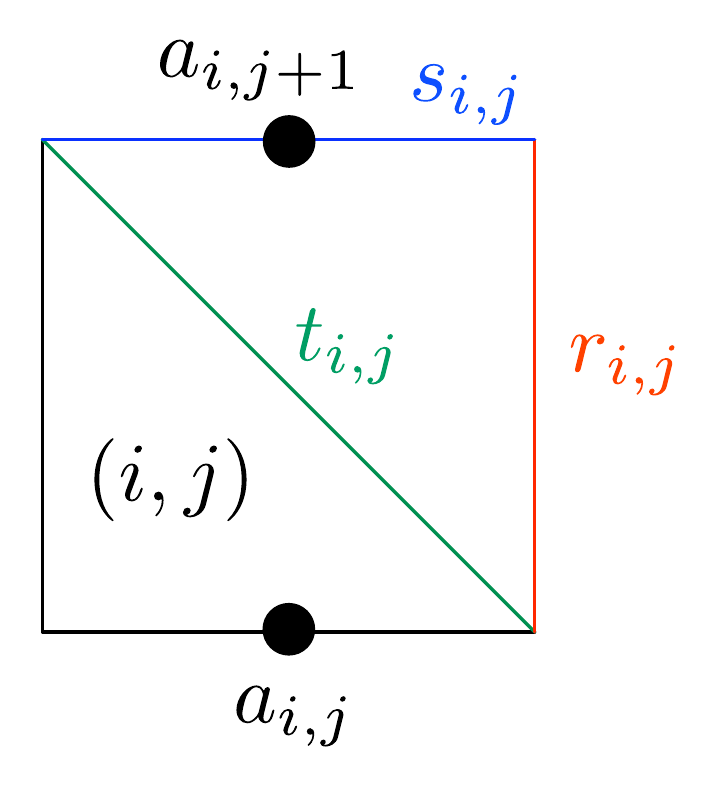}
\caption{\footnotesize{Labeling curve classes, and introducing $a$-parameters.}}
\label{fiducial_labeling}
\end{figure}

We have indicated a basis of $H_2(\CYX_0,\IZ)$ in figure \ref{fiducial_geometry}. Applying the labeling scheme introduced in figure \ref{fiducial_labeling}, the curve classes of our geometry are expressed in this basis as follows,
\ba
r_{i,j} &=& r_i + \sum_{k=1}^j (t_{i+1,k-1} - t_{i,k}) \\
s_{i,j} &=& s_j + \sum_{k=1}^i ( t_{k-1,j+1} - t_{k,j}) \,.
\ea
It proves convenient to express these classes as differences of what we will refer to as $a$-parameters \cite{part_1}, defined via
\beq
 t_{i,j}=a_{i,j}-a_{i,j+1} \quad\,, \quad \quad r_{i,j} = a_{i,j+1} - a_{i+1,j} \,.
\eeq

\subsection{The mirror of the fiducial geometry}
\label{secmirrorcurve}

The Hori-Vafa prescription \cite{HoriVafa} allows us to assign a mirror curve to a toric Calabi-Yau manifold. Each torically invariant divisor, corresponding to a 1-cone $\rho \in \Sigma(1)$, is mapped to a $\IC^*$ variable $e^{-Y_\rho}$. These are constrained by the equation
\beq
\sum_{\rho \in \Sigma(1)} e^{-Y_\rho} = 0 \,.
\eeq
Relations between the 1-cones, as captured by the lattice $\Lambda_h$ introduced in section (2.1) of \cite{part_1}, map to relations between these variables: for $\sigma \in \Sigma(2)$,
\beqn  \label{y_relations}
\sum_{\rho \in \Sigma(1)} \lambda_\rho(\sigma) Y_\rho = W_\sigma \,.
\eeqn
The $W_\sigma$ are complex structure parameters of the mirror geometry, related to the K\"ahler parameters $w_\sigma= r_{i,j}, s_{i,j}, \ldots$ introduced in the previous subsection via the mirror map, as we will explain in the next subsection.

The Hori-Vafa prescription gives rise to the following mirror curve $\curve_{\CYX_0}$ of our fiducial geometry $\CYX_0$, 
\begin{equation}
\sum_{i=0}^{n+1} \sum_{j=0}^{m+1} x_{i,j}  = 0  \,.  \label{HV}
\end{equation}
We have here labeled the 1-cones by coordinates $(i,j)$, beginning with $(0,0)$ for the cone $(0,0,1)$ in the bottom left corner of box $(0,0)$ as labeled in figure \ref{fiducial_geometry}, and introduced the notation
\beq
x_{i,j} = e^{-Y_{i,j}} \,.
\eeq
Eliminating dependent variables by invoking (\ref{y_relations}) yields an equation of the form
\beqn \label{mirror_curve}
\sum_{i=0}^{n+1} \sum_{j=0}^{m+1} c_{i,j} z_{i,j} = 0 \,.
\eeqn
Here, $$z_{i,j} = x_0^{1-i-j} x_1^i x_2^j \,,$$ where we have defined 
\beq 
x_0=x_{0,0}\,,\,\, x_1=x_{1,0}\,,\,\, x_2=x_{0,1}\,.
\eeq
$(x_0:x_1:x_2)$ define homogeneous coordinates on $\IC \IP^2$. The form of the equation is independent of the choice of triangulation of the toric diagram. What does depend on this choice are the coefficients $c_{i,j}$. It is not hard to write these down for the fiducial geometry $\CYX_0$ with the choice of basis for $H_2(\CYX_0,\IZ)$ indicated in figure \ref{fiducial_geometry}. Explicitly, the relations between the coordinates of the mirror curve (\ref{HV}) are
\begin{equation}
x_{i,0} = \frac{x_{i-1,0} x_{i-1,1}}{x_{i-2,1}} e^{R_{i-2}} \,,\quad x_{0,j}= \frac{x_{0,j-1} x_{1,j-1}}{x_{1,j-2}} e^{S_{j-2}} \,, \quad x_{i,j}=\frac{x_{i-1,j} x_{i,j-1}}{x_{i-1,j-1}} e^{T_{i-1,j-1}} \,.  \label{rel_bw_mirror_variables}
\end{equation}
Solving in terms of $x_0,x_1,x_2$ yields the coefficients $c_{0,0}=c_{0,1}=c_{1,0}=1$, 
\begin{align*}
c_{i,0} &=\exp\left[\sum_{k=1}^{i-1} (i-k) (R_{k-1} + T_{k-1,0})\right] \,, \\
c_{0,j} &=\exp\left[\sum_{l=1}^{j-1} (j-l) (S_{l-1} + T_{0,l-1})\right] \,,\\
\intertext{and for $i,j >0$}
c_{i,j} &= \exp\left[(i+j-1) T_{0,0} + \sum_{k=1}^{i-1} (i-k)(R_{k-1}+T_{k,0}) + \sum_{l=1}^{j-1} (j-l)(S_{l-1}+T_{0,l}) +  \sum_{k=1}^{i-1} \sum_{l=1}^{j-1}  T_{k,l}\right]  \,.
\end{align*}
Note that the number of coefficients $c_{i,j}$, up to an overall rescaling, is equal to the number of independent curve classes $r_i$, $s_j$, $t_{i,j}$.
 
\begin{figure}[h]
 \centering
  \includegraphics[width=6cm]{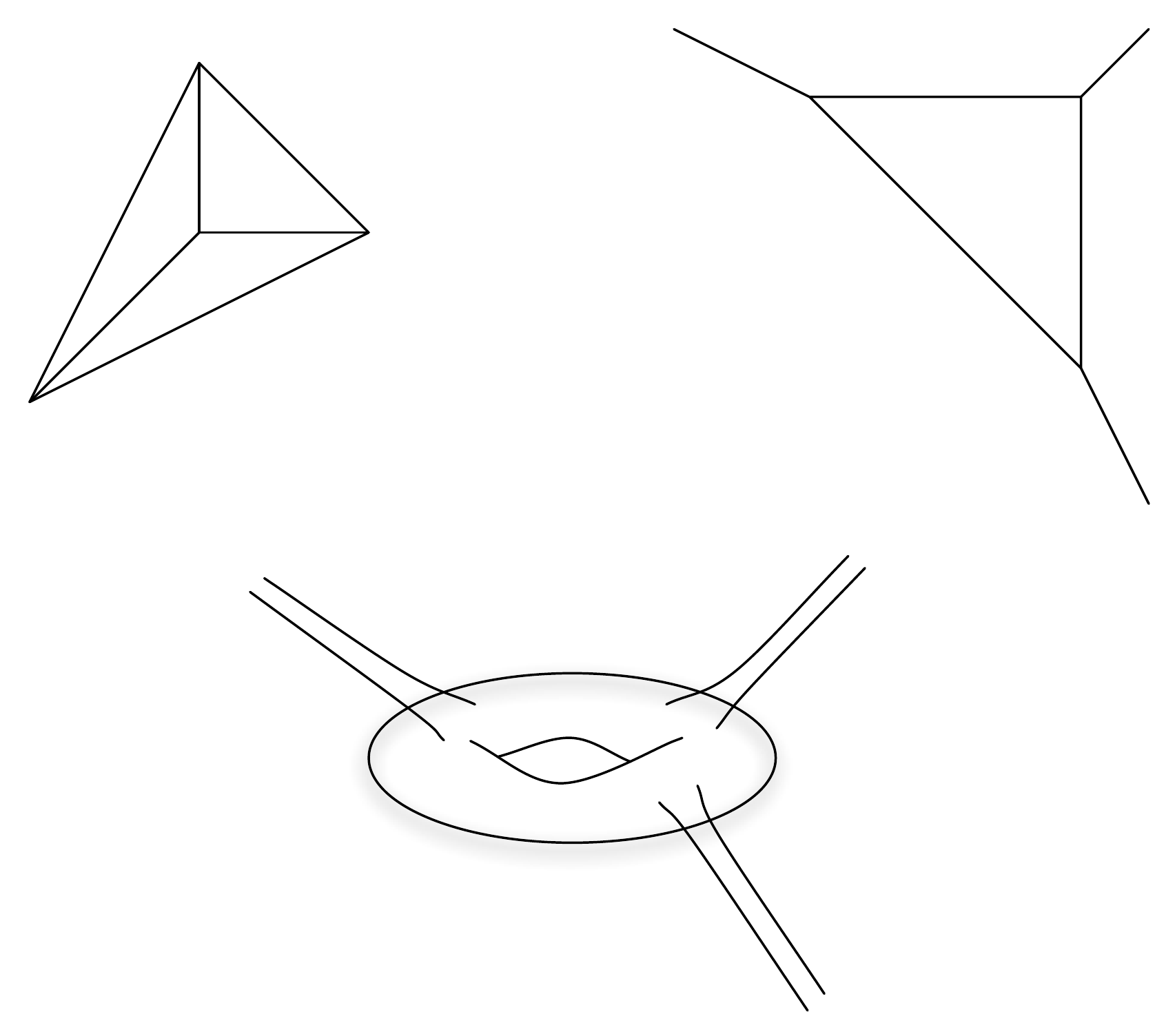}
 \caption{\footnotesize{Example of the thickening prescription: depicted are the fan for $\cO(-3) \rightarrow \IP^2$, the corresponding web diagram, and the mirror curve obtained via the thickening prescription.}}
 \label{thickening}
\end{figure}

In \cite{AKV}, the thickening prescription was put forth for determining the genus and number of punctures of the mirror curve: one is to thicken the web diagram of the original geometry to obtain the Riemann surface of the mirror geometry. The procedure is illustrated in figure \ref{thickening}. We will now verify this procedure by studying the curve (\ref{mirror_curve}) explicitly.

Let's consider the curve (\ref{mirror_curve}) for a single strip (i.e. $n=0$) of length $m+1$,
\begin{multline}
x_0^{m+2} + x_0^{m+1} x_1 + x_0^{m+1} x_2 + c_{1,1} \,x_0^{m} x_1 x_2 + c_{2,0}\, x_0^{m} x_1^2 + c_{2,1} \,x_0^{m-1} x_1^2 x_2 + c_{3,0} \,x_0^{m-1} x_1^3 + \ldots \\+ c_{m+1,0} \,x_0 x_1^{m+1} + c_{m+1,1} \,x_1^{m+1} x_2  = 0 \,. \label{curveonstrip}
\end{multline}
Note that the equation is of degree $m+2$, but the point $(0:0:1)$ is
an $m+1$-tuple point. By choosing the coefficients to be generic, we
can arrange for this singular point to be ordinary. The genus formula
then yields $$ {\frak g}= \frac{(d-1)(d-2)}{2} - \frac{m (m+1)}{2} = 0 \,.$$ In terms of the physical variables $Y_i$, any point on the curve with a vanishing homogeneous coordinate corresponds to a puncture. The punctures on the curve (\ref{curveonstrip}) thus lie at 
\begin{align*}
(0:0:1) \quad &: \quad m+1 \\
(0:1:0) \quad &: \quad 1 \\
(1:x_1^i:0) \quad&: \quad m+1 \\
(1:0:-1) \quad&: \quad 1 \,,
\end{align*}
where $x_1^i$, $i=1, \dots, m+1$, are the solutions of the equation
\ba
1 + x_1 + \sum_{j=1}^{m} d_i x_1^{i+1} &=&0 \,.
\ea
Note that we reproduce the $2m+4$ punctures expected from the thickening prescription of the toric diagram.

For the general case parametrized by $(m,n)$, the degree of the curve is $d=m+n+2$, and we have an ordinary $m+1$-tuple point at $(0:0:1)$ and an ordinary $n+1$-tuple point at $(0:1:0)$. The genus formula now yields $$g=\frac{(m+n)(m+n+1)}{2}-\frac{m(m+1)}{2}-\frac{n(n+1)}{2} = m n\,.$$ The punctures lie at
\begin{align*}
(0:0:1) \quad &: \quad m+1 \\
(0:1:0) \quad &: \quad n+1 \\
(1:x_1^i:0) \quad&: \quad m+1 \\
(1:0:x_2^j) \quad&: \quad n+1  \,,
\end{align*}
with $x_1^i$ the roots of $\sum_{i=0}^{m+1} c_{i,0} x_1^i = 0$ and $x_2^j$ the roots of $\sum_{j=0}^{n+1} c_{0,j} x_2^j = 0$. Again, we see that we reproduce the thickening prescription.

\subsection{The mirror map}
Above, we have distinguished between K\"ahler ($A$-model) parameters $w_\sigma$ and complex structure ($B$-model) parameters $W_\sigma$. At large radius/complex structure, these are identified between mirror pairs, but this identification is corrected by the so-called mirror map,\footnote{One could take exception to this nomenclature, arguing that the parameters $W_\sigma$ are the geometric parameters on both sides of the mirror, and refer to the $w_\sigma$ as the instanton or quantum corrected parameters. In such conventions, the curve classes in the various toric diagrams should be labeled by upper case letters.}
\beqn
W_\sigma = w_\sigma + \cO (e^{-w_\sigma}) \,. \label{mmap}
\eeqn
The exponentials of the parameters $W_\sigma$ appear as coefficients in the equation defining the mirror curve. They are global coordinates on the complex structure moduli space of the mirror curve. To compare expressions obtained in the A-model to those obtained in the B-model, all expressions are conventionally expressed in terms of flat coordinates $w_\sigma$. On the A-model side, these coordinates enter (in exponentiated form denoted generically as $Q_{\alpha,\beta}$ below) in the definition of the topological vertex. On the B-model side, they arise as the appropriate periods of a meromorphic one-form $\lambda$, defined in terms of the affine variables $x=\frac{x_1}{x_0}$, $y=\frac{x_2}{x_0}$ in the patch $x_0 \neq 0$ of the curve (\ref{mirror_curve}) as
\beq
\lambda = \log y \frac{dx}{x} \,.
\eeq
By calculating these periods as a function of the coefficients defining the mirror curve, we obtain the mirror map (\ref{mmap}).

The coordinates $w_\sigma$ are not globally defined functions on the complex structure moduli space. In the slightly clearer compact setting, this is due to the fact that the symplectic basis $\{\alpha_A,\beta^A\}$ of $H^3(\CYX,\IZ)$ in which we expand $\Omega$ (the compact analogue of the meromorphic 1-form $\lambda$ introduced above) such that the coefficients of $\alpha_A$ furnish our (local) coordinate system of the complex structure moduli space, undergo monodromy when transported around a singularity in moduli space.\footnote{Note that the symplectic basis makes no reference to complex structure, one might hence be led to believe that a global choice (i.e. one valid for any choice of complex structure) should be possible. This is not so. We consider the family $\pi: \cX \rightarrow \cS$, with $\cS$ the complex structure moduli space. The fiber over each point $w\in \cS$, $\pi^{-1}w = X_w$, is the Calabi-Yau manifold with the respective complex structure. $H^n(X_w,\IC)$ fit together to form a vector bundle $\cF_0$ over $\cS$, with a canonical flat connection, the Gauss-Manin connection. Using this connection, we can parallel transport a symplectic basis of $H^3(X_w,\IC)$ along a curve in $\cS$. As $\cS$ is not generically simply connected (due to the existence of degeneration points of the geometry), this transport may exhibit monodromy. Note that $\Omega$ can be defined as the section of a sheaf in the Hodge filtration of $H^3$ which extends to the singular divisor, hence is single valued. The monodromy in our choice of flat coordinates is therefore entirely due to the choice of symplectic basis.} A good choice of coordinates in the vicinity of a singular divisor $D$ hence involves a choice of basis forms that are invariant under monodromy around that divisor.

\section{Our matrix model} \label{s_our_matrix_model}

We derived a chain of matrices matrix model that reproduces the topological string partition function on $\CYX_0$ in \cite{part_1}. For $\CYX_0$ of size $(n+1)\times (m+1)$, as depicted in figure \ref{fiducial_geometry}, it is given by
\ban 
{Z}_{\rm MM}(\vec Q,g_s,\vec\alpha_{m+1},\vec\alpha_0^T)
&=& \Delta(X(\vec \alpha_{m+1}))\,\, \Delta(X(\vec \alpha_0)) \,\, 
\prod_{i=0}^{m+1} \int_{H_N(\Gamma_i)} dM_i \,
 \prod_{i=1}^{m+1}\int_{H_N({\mathbb R}_+)}\,dR_i \nn \\
&& \prod_{i=1}^{m} e^{{-1\over g_s}\,\tr \left[ V_{\vec a_i}(M_i)-V_{\vec a_{i-1}}(M_i) \right]
} \,\,\,
 \prod_{i=1}^{m} e^{{-1\over g_s}\,\tr \left[V_{\vec a_{i-1}}(M_{i-1})-V_{\vec a_{i}}(M_{i-1}) \right]
} \nn \\
&& \prod_{i=1}^{m+1} e^{{1\over g_s} \tr (M_i-M_{i-1})R_i} \,\,\,
 \prod_{i=1}^{m} e^{(S_i+{i\pi\over g_s})\,\tr\, \ln M_i}\,  \nn\\
&& e^{\tr \ln f_{0}(M_0)}\,\,e^{\tr \ln f_{m+1}(M_{m+1})}\,\, \prod_{i=1}^{m} e^{\tr \ln f_{i}(M_i)} \,. \label{m_integral}
\ean
We give the explicit expressions for the various functions entering in this definition in appendix \ref{our_matrix_model}. Here, we briefly explain some of its general features. 

The matrix model (\ref{m_integral}) is designed to reproduce the topological string partition function on the toric Calabi-Yau manifold $\CYX_0$ as computed using the topological vertex \cite{AKMV}. Recall that in this formalism, the dual web diagram to the toric diagram underlying the geometry is decomposed into trivalent vertices. Each such vertex contributes a factor $C(\alpha_i,\alpha_j,\alpha_k$) \cite{AKMV}, where $\alpha_i$ denote Young tableaux (partitions) of arbitrary size, one associated to each leg of the vertex. Legs of different vertices are glued by matching these Young tableaux and summing over them with appropriate weight.

Aside from the coupling constant $g_s$ and K\"ahler parameters of the geometry, denoted collectively as $\vec Q$, the matrix model (\ref{m_integral}) depends on partitions $\vec\alpha_0$, $\vec\alpha_{m+1}$ associated to the outer legs of the web diagram, which we choose to be trivial in this paper. The two classes of integrals $dR_i$ and $dM_i$ correspond to the two steps in which the topological string partition function on the fiducial geometry $\CYX_0$ can be evaluated: First, the geometry can be decomposed into $m+1$ horizontal strips, with partitions $\alpha_{j,i+1}$ and $\alpha_{j,i}$ associated to the upper and lower outer legs of the associated strip web diagram. $j = 0, \ldots, n$ counts the boxes in figure \ref{fiducial_geometry} in the horizontal direction, $i= 0, \ldots, m+1$ is essentially the strip index. Each such strip has a $dR_i$ integration associated to it. The partition function on such strips was calculated in \cite{IqbalKashaniPoor}. Following \cite{KlemmSulkowski}, we introduce two matrices $M_i$, $M_{i+1}$ per strip. Their eigenvalues encode the partitions $\alpha_{j,i}$ and $\alpha_{j,i+1}$ for all $j$. To work with finite size matrices, we introduce a cut-off $d$ on the number of rows of the Young tableaux we sum over. As we argue in section \ref{arctic}, our matrix model depends on $d$ only non-perturbatively. The strip partition function is essentially given by the Cauchy determinant of the two matrices $M_i$, $M_{i+1}$ \cite{part_1}, and the $dR_i$ integrals are the associated Laplace transforms. Gluing the strips together involves summing over the partitions $\alpha_{j,i}$. This step is implemented by the $dM_i$ integrations. To obtain a discrete sum over partitions from integration, we introduce functions $f_i(M_i)$ with integrally spaced poles. Integrating $M_i$ along appropriate contours then yields the sum over partitions as a sum over residues, the potentials $V_{\vec{a}_i}$ chosen to provide the proper weight per partition.

\section{Generalities on solving matrix models}  \label{generalities_on_mm}

\subsection{Introduction to the topological expansion of chain of matrices}

Chain of matrices matrix models have been extensively studied (see Mehta's book \cite{MehtaBook} and the review article \cite{DFGZJ}), and the computation of their topological expansion was performed recently in \cite{Eynchain, EPrats}.

\smallskip
The solution provided in \cite{EPrats} is based on the computation of the spectral curve $\spcurveMM$ of the matrix model. In \cite{EPrats, Eynchain}, only the case of potentials whose derivatives are rational functions is considered, and 
similarly to the one matrix model, the planar\footnote{\label{planar}For matrix models with $N$-independent polynomial potentials whose $g_s$ dependence is given by an overall prefactor, the planar limit coincides with the large $N$ limit, but this correspondence can fail if the potential or the integration contours have a non-trivial $N$ or $g_s$ dependence. The planar limit is defined by keeping only planar graphs in the Feynman graph perturbative expansion around an extremum of the potential. However, it is helpful to have in mind the intuitive picture that the planar limit is similar to a large $N$ limit.} expectation value of the resolvent of the first matrix of the chain is shown to satisfy an algebraic equation. The spectral curve is defined to be the solution locus of this equation. 
A general recipe is provided in \cite{Eynchain, EPrats} to obtain the spectral curve from algebraic equations and analyticity properties related to rational potentials and integration contours. Here, our potentials contain logs of $g$-functions. As they are not rational, we will have to present a slight extension of the recipe of \cite{EPrats} in section \ref{secspcurvegenchain}. This extension from rational potentials to analytical potentials, although not published, is straightforward, and  the derivation of these results will appear in \cite{toappear}.
In some sense, the derivative of $\ln g(x)$ can be viewed as a rational function with an infinite number of simple poles, i.e. as a limit of a rational function. More precisely, as an expansion in powers of $q$, to each order, it is a rational function. Since the spectral curve can be described by local properties, independent of the number of poles, one can take the limit of the recipe of \cite{Eynchain, EPrats}.
This is what we shall do in section \ref{secspcurvegenchain} below.

\smallskip
Having found the spectral curve $\spcurveMM$ of the matrix model, we will compute its symplectic invariants
$$
F_g(\spcurveMM)\, ,\quad g=0,1,2,3,\dots
$$
Symplectic invariants $F_g(\spcurve)$ can be computed for any analytical plane curve $\spcurve$, and thus in particular for $\spcurve=\spcurveMM$. For a general $\spcurve$ they were first introduced in \cite{EOFg}, as a generalization of the solution of matrix models loop equations of \cite{eynloop1mat}.
Their definition is algebraic and involves computation of residues at branch points of $\spcurve$.
We recall the definition below in section \ref{secdefspinv}.

\subsection{Definition of the general chain of matrices}

We consider chain of matrices matrix models of the form
\beqn\label{Zgeneralchain}
Z = \int_{{\cal E}} dM_1\dots dM_L\,\, e^{-{1\over g_s}\Tr \sum_{i=1}^L V_i(M_i)} \,\, e^{{1\over g_s}\Tr \sum_{i=1}^{L-1}  c_i M_{i} M_{i+1}} \,.
\eeqn
Note that aside from the potentials $V_i(M_i)$, the only interactions are between nearest neighbors, whence the name ``chain of matrices.'' Chain of matrices matrix models can be solved when the interaction terms between different matrices are of the form $\Tr M_{i} M_{i+1}$, as is the case here.

${\cal E}$ can be any ensemble of $L$ normal matrices of size $N\times N$, i.e. a submanifold of $\mathbb C^{LN^2}$ of real dimension $LN^2$, such that the integral is convergent. ${\cal E}$ can be many things; for a chain of matrices model, it is characterized by the contours on which eigenvalues of the various normal matrices are integrated (see \cite{MarcoPaths} for the 2-matrix model case). 
For (\ref{Zgeneralchain}) to have a topological expansion, ${\cal E}$ must be a so-called steepest descent ensemble (see \cite{EOreview}, section 5.5).
For a generic ensemble ${\cal E}$ which would not be steepest descent, $\ln Z$ would be an oscillating function of $1/g_s$, and no small $g_s$ expansion would exist, see \cite{BDE}.

The matrix model introduced in \cite{part_1} and reproduced in section \ref{s_our_matrix_model} was defined to reproduce the topological string partition function, which is defined as a formal series in $g_s$, and therefore has a topological expansion by construction.

An ensemble ${\cal E}$ is characterized by filling fractions $n_{j,i}$,
\beqn
{\cal E} = \prod_{i=1}^L {\cal E}_i
\quad , \quad
{\cal E}_i  = H_N(\gamma_{1,i}^{n_{1,i}} \times  \gamma_{2,i}^{n_{2,i}} \times \dots \times \gamma_{k_i,i}^{n_{k_i,i}}) \,,   \label{matrix_ensemble}
\eeqn
where $H_N(\gamma_1^{n_1}\times \dots\times \gamma_k^{n_k})$ is the set of normal matrices with $n_1$ eigenvalues on path $\gamma_1$, $n_2$ eigenvalues on path $\gamma_2$, $\dots$, $n_k$ eigenvalues on path $\gamma_k$.

As the filling fractions $n_{j,i}$ must satisfy the relation
\beq
\sum_{j=1}^{k_i} n_{j,i} = N
\eeq
for all $i$, only $\sum_i (k_i-1)$ of them are independent.

We also allow some paths $\gamma_{j,i}$ to have endpoints where $e^{-\Tr \sum_{i=1}^L (V_i(M_i) - M_{i} M_{i+1})}\neq 0$ -- indeed, in our matrix model, the matrices $R_i$ are integrated on $H(\mathbb R_+^N)$.

\subsubsection{The resolvent}\label{seceigenvalueinterp}
The spectral curve encodes all $W_i(x)$, the planar limits (see footnote \ref{planar}) of the resolvents of the matrices $M_i$,
\beq
W_i(x) = g_s\,\left< \tr {1\over x-M_i}\right>_{planar} \,,
\eeq
see equation (\ref{resolvent}) below. The respective $W_i$ can be expressed as the Stieljes transform
\beq
W_i(x) = \int {\rho_i(x')dx'\over x-x'} 
\eeq
of the planar expectation value of the eigenvalue density $\rho_i(x)$ of the matrix $M_i$,
\beq
\rho_i(x) =g_s\,\left< \tr \delta(x-M_i)\right>_{planar} \,.
\eeq

By general properties of Stieljes transforms, singularities of $W_i(x)$ coincide with the support of the distribution $\rho_i(x)dx$:
\begin{itemize}
\item Simple poles of $W_i(x)$ correspond to delta distributions i.e. isolated eigenvalues.

\item Multiple poles correspond to higher derivatives of delta distributions.

\item Cuts correspond to finite densities, the density being the discontinuity of $W_i(x)$ along the cut,
\beqn  \label{density_as_discontinuity}
\rho_i(x) = {1\over 2i\pi}\,(W_i(x-i0)-W_i(x+i0)) \,.
\eeqn
\end{itemize}

In particular, cuts emerging from algebraic singularities (generically square root singularities) correspond to densities vanishing algebraically (generically as square roots) at the endpoints of the cut. Cuts emerging from logarithmic singularities correspond to constant densities.

\subsubsection{The spectral curve of the general chain of matrices}
\label{secspcurvegenchain}

When all $V'_i$ are rational, the spectral curve was found in \cite{Eynchain, EPrats}, and it is algebraic.
We present here a generalization of this result to more general potentials. The derivations of these results will appear in \cite{toappear}.

The spectral curve can be obtained by the following procedure:
\medskip

\begin{enumerate}
\item \label{genus_ss} Consider a compact Riemann surface $\curve$ of genus
\beq
{\mathfrak g}= \sum_{i=1}^{L} (k_{i}-1)   \,,
\eeq
where $k_i$ denotes the number of cuts of the $i$-th matrix, as implicitly defined in (\ref{matrix_ensemble}).

\item \label{projections}
Look for $L+2$ functions on $\curve$, 
\beq
x_0(z),\, x_1(z),\, x_2(z),\, \dots , \, x_{L}(z),\, x_{L+1}(z): \curve \rightarrow \IC \IP^1 \,.
\eeq
The $x_i$ are to be holomorphic away from points $z \in \curve$ at which $V'_{i-1}(x_{i-1})$ or $V'_{i+1}(x_{i+1})$ become singular, and satisfy the functional relations
\beqn\label{eqspcurvechainxxV}
c_{i-1}x_{i-1}(z)+c_i x_{i+1}(z) = V'_i(x_i(z)) \,.
\eeqn
Recall that the $c_i$ are the coefficients of the interaction potentials in (\ref{Zgeneralchain}). We have set $c_0=c_{L}=1$.

For each $i=1,\dots,L$, the Riemann surface $\curve$ can be realized as a branched covering of $\mathbb C\IP^1$ by the projection $x_i:\curve\to \mathbb C\IP^1$. A choice of branched covering is not unique: the choice consists in the set of cuts connecting branch points (recall that these are points at which $dx_i(z)=0$). We will determine an appropriate covering below in step \ref{sings}.

\item \label{hard_edges} If some path $\gamma_{j,i}$ has an endpoint $a$ (called ``hard edge'' in the matrix model literature, see \cite{Eynhardedge}), then choose a pre-image $a_i \in x_i^{-1}(a)$ and require
\beq
dx_i(a_i)=0
\quad {\rm and} \quad
x_{i-1}(z)\,\,{\rm has\, a \,simple\,pole\, at\,} z=a_i.
\eeq
The topological recursion is proved in \cite{EPrats} without hard edges, but it is not difficult to see, by mixing the results of \cite{Eynhardedge}, \cite{ChekhovHardedges} and \cite{EPrats}, that the topological recursion continues to hold in the presence of hard edges. The proof will appear in a forthcoming publication \cite{toappear}. Here, we shall assume that it holds.

\item \label{sings} Choose some contours $\widehat{\cal A}_{j,i}$, $j=1,\dots,k_i$ in $\mathbb C \IP^1$, such that  each $\widehat{\cal A}_{j,i}$ surrounds all points of the contour $\gamma_{j,i}$ (related to the matrix ensemble ${\cal E}_i$ defined in (\ref{matrix_ensemble})) in the clockwise direction and no other contour $\gamma_{j',i}$. 
For $x\in \mathbb C \IP^1$ not enclosed in the contours $\widehat{{\cal A}}_{j,i}$, $j=1,\dots,k_i,$ and given a connected component ${{\cal A}_{j,i}}$ of the pre-image of the contour $\widehat{{\cal A}}_{j,i}$ under $x_i$,
\beq
{\cal A}_{j,i} \subset x_i^{-1}(\widehat{{\cal A}}_{j,i}) \,,
\eeq
 define the function
\beqn  \label{resolvent}
W_i(x) = {c_{i-1}\over 2i\pi}\sum_{j=1}^{k_i}\,\oint_{{\cal A}_{j,i}}\,\, {x_{i-1}(z)\,dx_i(z)\over x-x_i(z)}.
\eeqn

Generalizing \cite{Eynchain} to non-polynomial potentials, we claim that a choice of 
${\cal A}_{j,i}$ exists such that $W_i(x)$ is the planar limit of the resolvent of the matrix $M_i$. In the following, it is this choice that will be referred to as ${\cal A}_{j,i}$.

Notice that not all ${\cal A}_{j,i}$ will be homologically independent on $\curve$. 
We require that we have $\mathfrak g=\sum_{i=1}^L (k_i-1)$ homologically independent ${\cal A}_{j,i}$'s, which coincides with the genus of $\curve$. As a condition on the choice of branched covering, we impose that ${\cal A}_{j,i}$ and $a_i$ lie on the same sheet of $x_i$. This condition, in our experience, uniquely fixes this choice. We will assume that this is the case. We refer to the sheet of $x_i$ containing ${\cal A}_{j,i}$ and $a_i$ as the physical sheet for $x_i$.

\item \label{filling_fraction} In accord with (\ref{density_as_discontinuity}), we consider the discontinuity of $W_i(x)$ along the $j$-th cut. It is given by
\ban \label{disc}
\Disc_j\, W_i(x) &=& \frac{1}{2\pi i} \left( W_i(x_+) - W_i(x_-) \right) \nn \\
&=& \frac{1}{2\pi i} c_{i-1} \,  \Disc_j\, x_{i-1} \,, \label{rho}
\ean
as we explain in figure \ref{discontinuity}.
\begin{figure}[h]
 \centering
  \includegraphics[width=5cm]{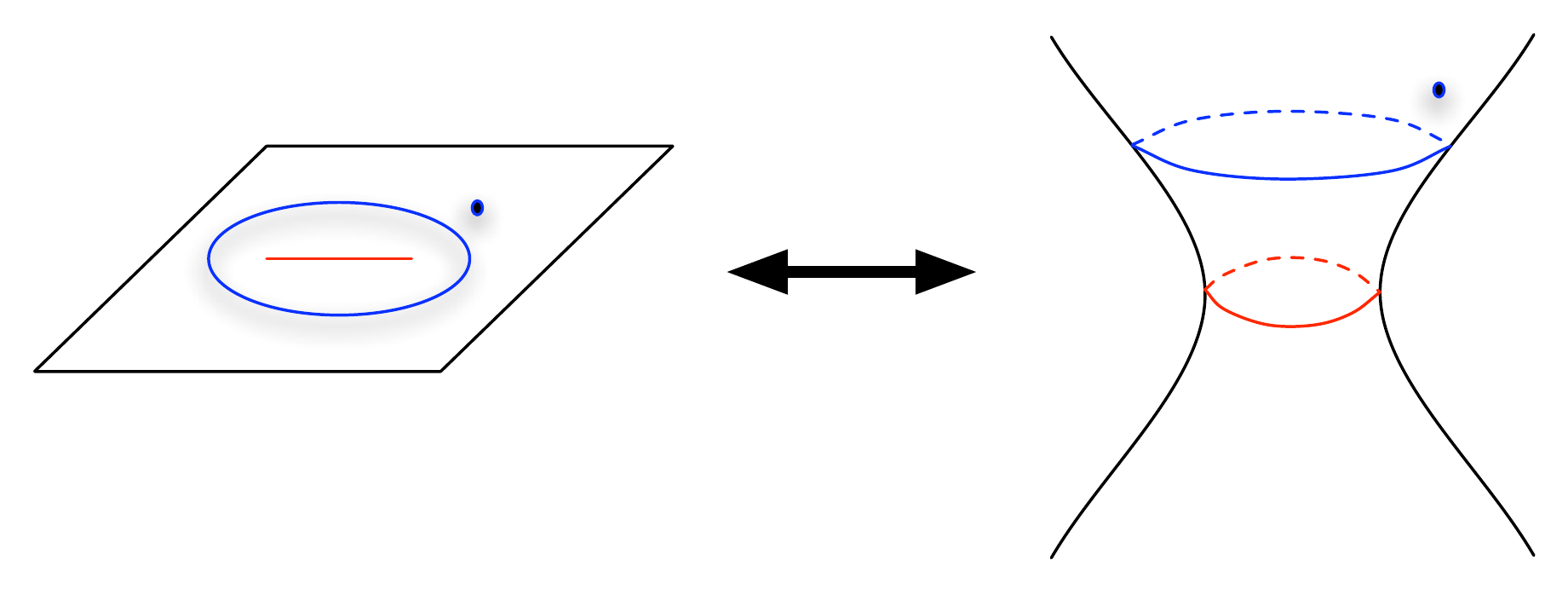}
 \caption{\footnotesize{The integration contour $\widehat{\cal A}_{j,i}$ on the $x$-plane, and its image ${\cal A}_{j,i}$ on $\curve$.}}
 \label{cut}
\end{figure}
\begin{figure}[h]
 \centering
  \includegraphics[width=16cm]{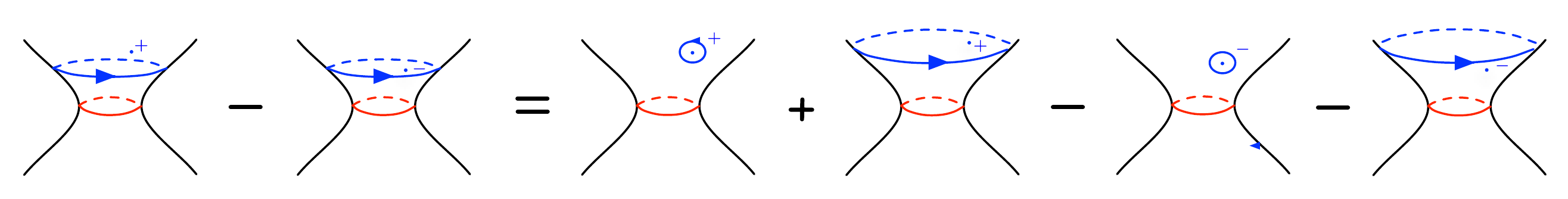}
 \caption{\footnotesize{The preimage of the points $x_+$ and $x_-$ of (\ref{disc}) are depicted as dots in the above diagram, $\widehat{{\cal A}}_{j,i}$ is given by the blue contour, and the preimage of the cut is drawn in red. To take the limit $x_+ \rightarrow x_-$, one must first shift the contours. The second and fourth term on the RHS of the above diagrammatic equation then cancel, yielding the RHS of (\ref{rho}).}}
 \label{discontinuity}
\end{figure}

The definition (\ref{matrix_ensemble}) of the matrix ensemble ${\cal E}_i$ is the condition that there are $n_{j,i}$ eigenvalues of $M_i$ on the contour $\gamma_{j,i}$, hence corresponds to imposing the filling fraction conditions 
\beq
\frac{1}{2\pi i}\oint_{{\cal A}_{j,i}} c_{i-1}x_{i-1} dx_i = g_s\,\, n_{j,i}
\eeq
for $i=1,\dots,L$, $j=1,\dots,k_i$.
\end{enumerate}

\bigskip

In our experience,  the conditions enumerated above have a unique solution and define a unique spectral curve. As emphasized in the introduction, a formal uniqueness proof is however still lacking.

The spectral curve is defined as the data of the Riemann surface $\curve$, and the two functions $x_1(z)$ and $x_2(z)$,
\beqn  \label{spectral_curve}
\encadremath{
\spcurve_{MM} = (\curve,x_1,x_2).
}\eeqn

\subsection{Symplectic invariants of a spectral curve}
\label{secdefspinv}

Once we have found the spectral curve $\spcurve_{MM}$ of our matrix model, we can compute the coefficients $F_g$ in the topological expansion of its partition function,
\beq
\ln Z= \sum_{g=0}^\infty g_s^{2g-2} F_g  \,,
\eeq
by computing the symplectic invariants of this curve,
\beq
F_g=F_g(\spcurve_{MM}) \,,
\eeq
 following \cite{EPrats}.
\smallskip

Let us recall the definition of these invariants for an arbitrary spectral curve $\spcurve$. 
\bigskip

Let $\spcurve=(\curve,x,y)$ be a spectral curve, comprised of the data of a Riemann surface $\curve$ and two functions $x(z),\, y(z): \curve \rightarrow \IC$, meromorphic on $\curve$ away from a finite set of points (we wish to allow logarithms).\footnote{In fact, the most general setting in which this formalism is valid has not yet been established. We will state it within the generality we need here i.e. we assume that $dx$ is meromorphic forms on $\curve$  (this allows $x$ and $y$ to have logarithms).} We will assume that $dx$ is a meromorphic form on all of $\curve$.

\subsubsection{Branchpoints}

\begin{figure}[h]
 \centering
  \includegraphics[width=6cm]{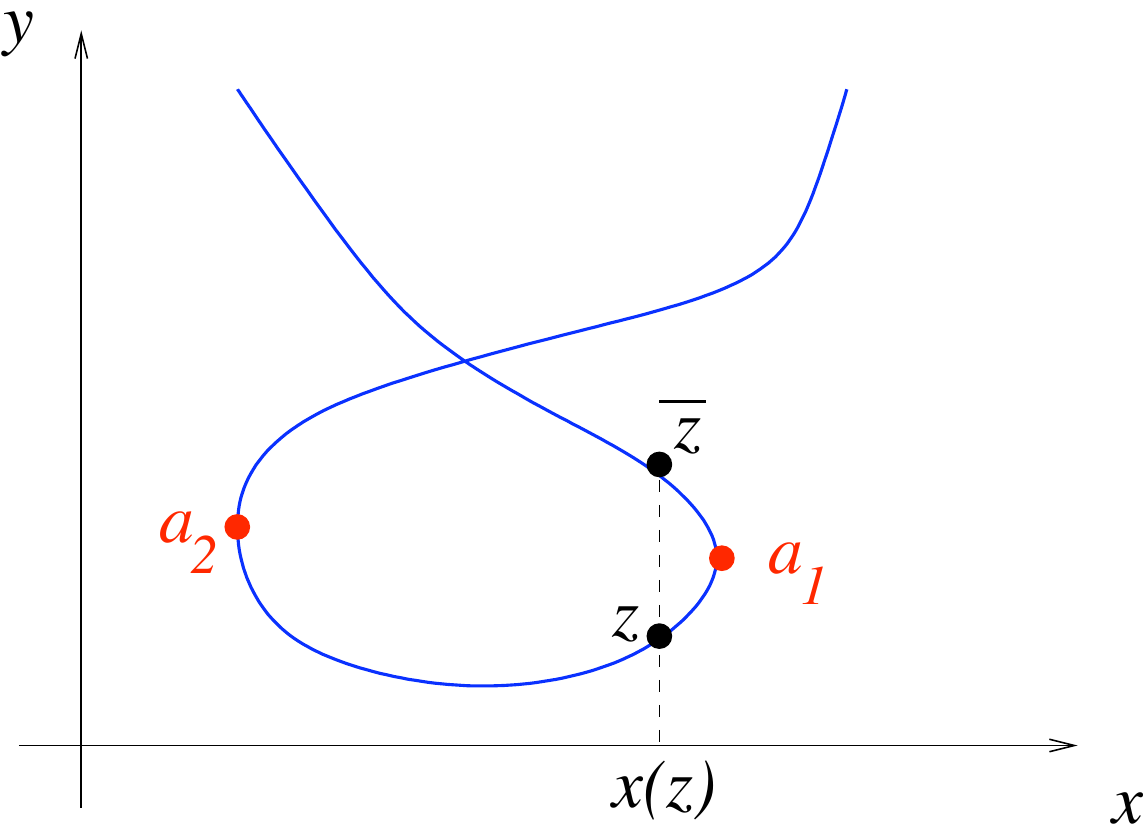}
 \caption{\footnotesize{At a regular branch point $a\in \curve$ of $x$, $y$ as a function of $x$ has a branchcut $y\sim y(a) + C \sqrt{x-x(a)}$. If $z$ is a point on one branch near $a$, we call $\bar z$ the conjugate point on the other branch; it has the same $x$ projection, $x(\bar z)=x(z)$. Notice that $\bar z$ is defined only locally near branch points. If we follow $z$ from $a_1$ to $a_2$, $\bar z$ may have to jump from one branch to another.
 }}
 \label{figspcurvebpzzbar}
\end{figure}

Let $a_i$ be the branch points of the function $x$,
\beq
dx(a_i)=0.
\eeq
We assume that all branch points are simple, i.e. that $dx$ has a simple zero at $a_i$. This implies that in the vicinity of $a_i$, the map $x$ is $2:1$. We introduce the notation $\bar z\neq z$ such that
\beq
x(\bar z)=x(z).
\eeq
$\bar z$ is called the conjugate point to $z$, and it is defined only in the vicinity of branch points, as depicted in figure \ref{figspcurvebpzzbar}.

We also require that the branch points of $x$ and $y$ do not coincide, such that $dy(a_i)\neq 0$ and $y(z)$ therefore has a square-root branchcut as a function of $x$ at $x(a_i)$. If $y$ is finite at $a_i$, its local behavior is hence given by
\beq
y(z) \sim y(a_i)  + C_i\sqrt{x(z)-x(a_i)} \,.
\eeq
If $a_i$ corresponds to a hard edge, we require $y$ to have a pole here. Its local behavior is hence given by 
\beq
y(z) \sim  {C_i\over \sqrt{x(z)-x(a_i)}} \,.
\eeq

\subsubsection{Bergman kernel}

On a curve $\curve$, there exists a unique symmetric 2-form $B(z_1,z_2)$ with a double pole on the diagonal $z_1=z_2$ and no other poles, with the following normalization on ${\cal A}$-cycles,
\beq
\oint_{z_2\in{\cal A}_{j,i}} B(z_1,z_2)=0.
\eeq
In any local coordinate near $z_1=z_2$, one has 
\beq
B(z_1,z_2) \sim {dz_1\,dz_2\over (z_1-z_2)^2} + {\rm regular}  \,.
\eeq 
$B$ is called the Bergman kernel of $\curve$, or the fundamental 2-form of the second kind \cite{Fay}.

\subsubsection{Recursion kernel}

We now define the recursion kernel $K$ as
\beq
K(z_0,z) = \frac{\int_{\bar z}^z B(z_0,z')}{2(y(\bar z)-y(z))\,dx(z)} \,.
\eeq
This kernel is a globally defined 1-form in the variable $z_0\in \curve$. In the variable $z$, it is the inverse of a 1-form (that means we have to multiply it with a quadratic differential before computing any integral with it); it is defined only locally near branch points of $x$, such that $K(z_0,\bar z)=K(z_0,z)$. At the branch points, it has simple poles,
\beq
K(z_0,z) \sim -{B(z_0,z)\over 2\,dx(z)\,dy(z)} + {\rm regular}.
\eeq

\subsubsection{Topological recursion}

Correlation forms $W_n^{(g)}(z_1, \ldots, z_n)$ (not to be confused with the resolvents $W_i(z)$ introduced above) are symmetric $n$-forms defined by
\beq
W_1^{(0)}(z) = -y(z)dx(z) \,,
\eeq
\beq
W_2^{(0)}(z_1,z_2) = B(z_1,z_2) \,,
\eeq
and then by recursion (we write collectively $J=\{z_1,\dots,z_n\}$),
\ba
W_{n+1}^{(g)}(z_0,J) 
&=& \sum_i \Res_{z\to a_i} K(z_0,z)\, \Big[ W_{n+2}^{(g-1)}(z,\bar z,J) \cr
&& + \sum_{h=0}^{g}\sum'_{I\subset J} W_{1+|I|}^{(h)}(z,I)\,W_{1+n-|I|}^{(g-h)}(\bar z,J\setminus I) \Big]
\ea
where $\sum'_I$ is the sum over all subsets of $J$, restricted to $(h,I)\neq (0,\emptyset)$ and $(h,I)\neq (g,J)$.

\medskip

Although it is not obvious from the definition, the forms $W_n^{(g)}$ are symmetric.
For $2-2g-n<0$, they are meromorphic $n$-forms with poles only at branch points. These poles are of degree at most $6g-4+2n$, and have vanishing residues.

For the one matrix model, the $W_n^{(g)}$ coincide with the $n$-point function of the trace of the resolvent at order $g$ in the topological expansion.

\subsubsection{Symplectic invariants}
\label{secdefspinvs}

Finally, for $g\geq 2$, we define the symplectic invariants $F_g$ (also denoted $W_0(g)$ in \cite{EOFg}) by
\beq
F_g(\spcurve)  = {1\over 2-2g}\, \sum_i \Res_{z\to a_i} \Phi(z)\,W_1^{(g)}(z) \,,
\eeq
where $\Phi$ is any function defined locally near branch points of $x$ such that $d\Phi=y dx$.

The definitions of $F_1$ and $F_0$ are more involved and we refer the reader to \cite{EOFg}. $F_0$ is called the prepotential, and $F_1$ is closely related to the determinant of the Laplacian on $\curve$ with metrics $|ydx|^2$, see \cite{KK, EKK}. 

\medskip

The $F_g(\spcurve)$'s depend only on the orbit of $\spcurve$ under the group of transformations generated by
\begin{itemize} 
\item[$\mathfrak{R}:$] $\spcurve\mapsto\td\spcurve=(\curve,x,y+R(x))$ where $R(x)$ is any rational function of $x$,
\item[$\mathfrak{F}:$] $\spcurve\mapsto\td\spcurve=(\curve,f(x),y/f'(x))$ where $f(x)$ is an analytical function of $x$, with $f'$ rational, such that $df=f'dx$ has the same number of zeroes as $dx$,
\item[$\mathfrak{S}:$] $\spcurve\mapsto\td\spcurve=(\curve,y,-x)$.
\end{itemize}
These transformations are symplectic, i.e. they leave $dx\wedge dy$ invariant.

\medskip

The symplectic invariants are homogeneous of degree $2-2g$,
\beqn\label{Fghomogeneous}
F_g(\curve,x,\lambda y) = \lambda^{2-2g}\,F_g(\curve,x,y).
\eeqn
In particular, they are invariant under the parity transformation  $F_g(\curve,x,- y) = F_g(\curve,x,y)$.

\section{The spectral curve for the topological string's matrix model} \label{the_spectral_curve}

Applying the procedure outlined in section \ref{secspcurvegenchain} to our matrix model, we will determine a spectral curve $\spcurve_{MM}(\CYX_0)$ in this section. \cite{EPrats} demonstrated that for a chain of matrices, we have
\beq
\ln {Z} = \sum_g g_s^{2g-2} F_g(\spcurveMM) \,,
\eeq
with $F_g$ the symplectic invariants of \cite{EOFg}. In our case, since we have engineered our matrix model to yield\footnote{As we have here reserved the notation $F_g$ for the symplectic invariants of our matrix model, we refer to the topological string free energies as $GW_g$.} $GW_g(\CYX_0)$ as its partition function, re-computing the partition function via the methods of \cite{EPrats} will yield
\beq
GW_g(\CYX_0) = F_g(\spcurveMM)  \,.
\eeq
This relation is already quite interesting, as it allows for explicit computation of the Gromov-Witten invariants. Our goal however will be to go further. We will argue that $\spcurveMM$ is symplectically equivalent to the mirror spectral curve $\spcurve_{\hat \CYX_0}$ of section \ref{secmirrorcurve},
\beq
\spcurveMM \sim \spcurve_{\CYX_0} \,.
\eeq
Since the $F_g$'s are symplectic invariants, this will imply the BKMP conjecture for $\CYX_0$, i.e.
\beq
GW_g(\CYX_0) = F_g(\spcurve_{\CYX_0}).
\eeq

\subsection{Applying the chain of matrices rules} \label{applying_the_rules}

We now apply the rules of section \ref{secspcurvegenchain} to the chain of matrices model introduced in section \ref{s_our_matrix_model}.

\medskip
\begin{itemize}
\item Recall that the integration ensembles for the matrices $M_0$ and $M_{m+1}$ are such that for each matrix, all eigenvalues are integrated on the same contour (\ref{outer_contours}). Hence, $k_0=k_{m+1}=1$, and the corresponding filling fractions are equal to $N$.
For $i=1,\dots, m$, the matrix $M_i$ is integrated on $H(\gamma_{0,i}^{d} \times  \gamma_{2,i}^{d} \times \dots \times \gamma_{n,i}^{d})$, where $\gamma_{j,i}$ is a contour which surrounds all points of the form $q^{a_{j,i}+\mathbb N}$. There are thus $k_i=n+1$ filling fractions, each equal to $d$.
The matrices $R_i$ are integrated on $H(\mathbb R_+^N)$. We denote the number of their cuts by $\tilde{k_i}$. Hence, $\td k_i=1$, with the respective filling fraction equal to $N$.

According to condition \ref{genus_ss} of section \ref{secspcurvegenchain}, the genus of the spectral curve $\curve$ is thus given by 
$$
\mathfrak g=\sum_{i=0}^{m+1} (k_i-1)+\sum_{i=1}^{m+1} (\td k_i-1)=nm  \,.
$$

\item Following condition \ref{projections} of section \ref{secspcurvegenchain}, we introduce functions $x_i(z)$, $i=0,\dots,m+1$, associated to the matrices $M_i$, and functions $y_i(z)$, $i=1,\dots,m+1$, associated to the matrices $R_i$, as well as two additional functions $y_0(z)$ and $y_{m+2}(z)$ at the ends of the chain.

They must satisfy the following requirements:
\begin{itemize}

\item Since there is no potential for the matrices $R_i$, equation (\ref{eqspcurvechainxxV}) implies that we have, for $i=1,\dots,m+1$,
\beq
x_{i}(z)-x_{i-1}(z) = 0  \,.
\eeq
We can hence suppress the index $i$ on these functions, $x(z)=x_i(z)$.

\item For $i=1,\dots,m$, equation (\ref{eqspcurvechainxxV}) gives
\beq
y_i(z)-y_{i+1}(z) 
=  2 V'_{\vec a_i}(x(z))-V'_{\vec a_{i+1}}(x(z))-V'_{\vec a_{i-1}}(x(z))   
 - g_s {f'_i(x(z))\over f_i(x(z))} -{g_s \, S_i+i\pi\over x(z)}
\eeq
and
\beq
y_0(z)-y_{1}(z) 
=  V'_{\vec a_0}(x(z))-V'_{\vec a_{1}}(x(z))   - g_s {f'_0(x(z))\over f_0(x(z))} \,,
\eeq
\beq
y_{m+1}(z) -y_{m+2}(z)
=  V'_{\vec a_{m+1}}(x(z))-V'_{\vec a_{m}}(x(z))  - g_s {f'_{m+1}(x(z))\over f_{m+1}(x(z))} \,.
\eeq

More explicitly, in terms of the function $$\psi_q(x)=x g'(x)/g(x)\,,$$ whose small $g_s$ expansion 
\ba
\psi_q(x) &=& -\,{1\over \ln q}\,\sum_{n=0}^\infty {(-1)^n\,B_n\over n!}\,(\ln q)^n\,\, \Li_{1-n}(1/x) \\
&=&  {1\over \ln{q}}\,\Big[
\ln{(1-{1\over x})} -{\ln q\over 2(x-1)}
-\sum_{n=1}^\infty {B_{2n}\over (2n)!}\,\, (\ln q)^{2n}\,\Li_{1-2n}(x) \Big] \,.
\ea
we worked out in appendix A of \cite{part_1}, we obtain
\ban
\lefteqn{x(z)(y_{i+1}(z)-y_{i}(z) )}  \nn \\
&=& i\pi+ g_s S_i - g_s \sum_j (2\psi_q(q^{a_{j,i}}/x(z)) - \psi_q(q^{a_{j,i+1}}/x(z))-\psi_q(q^{a_{j,i-1}}/x(z))) \nn \\
&& + g_s {x(z)f'_i(x(z))\over f_i(x(z))} \,,    \label{diff_ys}
\ean
as well as
\ba
x(z)(y_1(z)-y_{0}(z) )
&=& -g_s \sum_j \psi_q(q^{a_{j,0}}/x(z))+g_s\sum_j\psi_q(q^{a_{j,1}}/x(z))  \cr
&& -g_s\sum_j\sum_{k=0}^{d-1} \frac{x(z)}{x(z)-q^{a_{j,0}+k}} \,,
\ea
\ba
x(z)(y_{m+2}(z) -y_{m+1}(z))
&=&  -g_s \sum_j \psi_q(q^{a_{j,m+1}}/x(z))+g_s\sum_j\psi_q(q^{a_{j,m}}/x(z)) \cr
&& -g_s\sum_j\sum_{k=0}^{d-1} \frac{x(z)}{x(z)-q^{a_{j,m+1}+k}}
\ea
Note that we have explicitly used the fact that the partitions $\alpha_{j,m+1}$ and $\alpha_{j,0}$ are chosen to be trivial.\\

\item Since the integral over $R_i$ is over $H_N(\mathbb R_+)$, i.e. its eigenvalues are integrated on $\mathbb R_+$, the integration contour has an endpoint (hard edge) at $y_i=0$. Condition \ref{hard_edges} hence requires that at a pre-image $y_i^{-1}(0)$, which we will refer to as $\infty_i$, the following holds
\beq
y_i(\infty_i)=0\, , \quad dy_i(\infty_i)=0\, , \quad x(z){\rm \,\, has\,a \,simple\,pole\,at\,}z=\infty_i \,.
\eeq

Furthermore, introducing a local parameter $z$ in the neighborhood of $\infty_i$, the above translates into 
\beq
y_i(z) \sim z^2 \,, \quad x(z) \sim 1/z  \,.
\eeq
Hence, $\forall \,i=1,\dots,m+1$,
\beq
y_i \sim \cO(1/x^2) \,.
\eeq
\end{itemize}

\item
The relations (\ref{diff_ys}) imply that near $\infty_i$, we have
\beq
x (y_{j+1}-y_j) \mathop{{\sim}}_{z\to\infty_i} i\pi + g_s S_j+ g_s\sum_{l=0}^n (2a_{l,j}-a_{l,j+1}-a_{l,j-1}) + O(1/x) \,.
\eeq
In particular, it follows that $\infty_j\neq \infty_i$. Thus, all points $\{\infty_1,\dots,\infty_{m+1}\}\subset x^{-1}(\infty)$ are distinct , i.e. condition \ref{hard_edges} requires that $x^{-1}(\infty)$ have at least $m+1$ points. 

We will make the minimal assumption that $x^{-1}(\infty)$ has exactly
$m+1$ elements that are simple poles of $x$, and that $x$ has no further singularities, i.e. that $x$ is a meromorphic function of degree  $m+1$ on $\curve$. 

\item By condition \ref{filling_fraction}, since for $i=0,\dots,m+2$ there are $d$ eigenvalues of $M_i$ of the form $q^{a_{j,i}+\mathbb N}$ surrounded by the path $\widehat{\cal A}_{j,i}$ , we have the $(m+2)\times (n+1)$ filling fraction conditions
\beq
{1\over 2i\pi}\oint_{{\cal A}_{j,i}} y_i\,dx = d\, g_s \qquad \mathrm{for}\,\, i=0,\dots,m+1, \,\,\, j=0,\dots,n \,.
\eeq
\end{itemize}

\begin{figure}[!t]
 \centering
  \includegraphics[width=6cm]{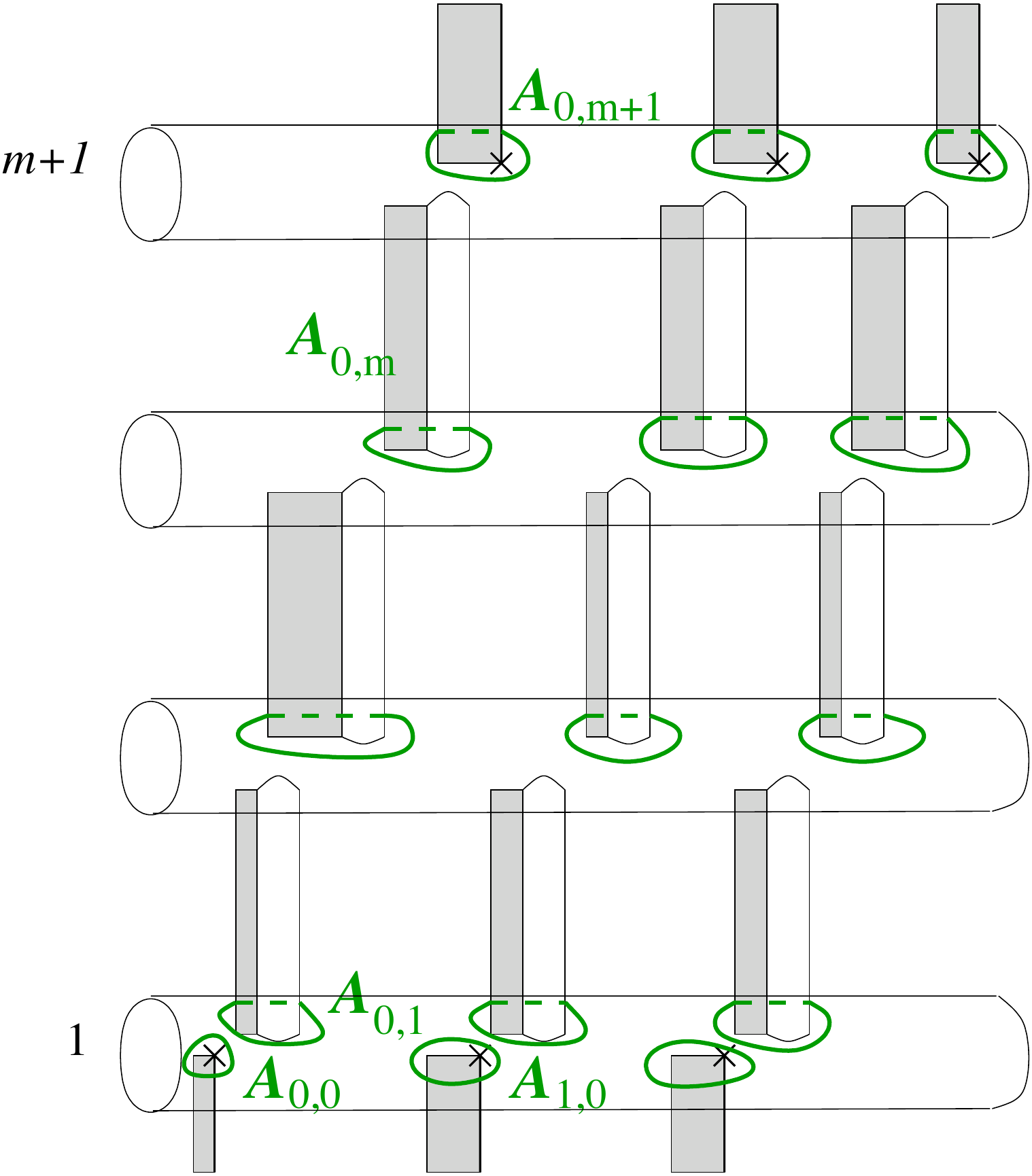}
 \caption{\footnotesize{The spectral curve of our matrix model can be represented as follows. The cover of $\IC \IP^1$ provided by $x$ has $m+1$ sheets. Instead of the projective plane of $x$, we represent the sheets of $\ln x$, which are cylinders.
 Cycles ${\cal A}_{j,i}$ appear in sheets $i-1$ and $i$. They enclose singularities of the resolvent $W_i$. Algebraic cuts are represented as vertical cylinders, and poles and log singularities are represented as grey strips. There is only one cycle ${\cal A}_{j,0}$ (which is in sheet $0$) and one ${\cal A}_{j,m+1}$ (in sheet $m$), and they enclose only poles or log singularities of $y_0$ resp. $y_{m+1}$.
 }}
 \label{figspcurve1}
\end{figure}

$x$ hence defines an $m+1$ sheeted cover of $\IC \IP^1$. Considering the function $\ln x$ instead, with singularities at $x=0$ and $x = \infty$, each sheet of this cover is mapped to a cylinder. We have depicted this covering in figure \ref{figspcurve1}, and indicated the singularities of $y_i$ on each sheet: algebraic cuts are represented by vertical cylinders, and poles and logarithmic cuts by grey strips.

In sheet $i$ we have represented some contours ${\cal A}_{j,i}$ whose image under the projection $x:\curve \to \mathbb C \IP^1$ surrounds all points of type $q^{a_{j,i}+\mathbb N}$. 

For $i=1,\dots,m$, the resolvent $W_i(x)$ of the $i^{\rm th}$ matrix $M_i$ is computed as a contour integral around the sum over $j$ of cycles ${\cal A}_{j,i}$ on sheet $i$,
\beq
W_i(x) = \sum_{j=0}^n\, {1\over 2i\pi}\oint_{{\cal A}_{j,i}}\, {y_i(z')dx(z')\over x-x(z')}  \,.
\eeq

Also, as argued in \cite{part_1}, the potentials of $M_0$ and $M_{m+1}$ are such that in fact the matrices $M_0$ and $M_{m+1}$ are frozen, and thus their resolvents contain only poles.
In terms of the functions $y_0$ and $y_{m+1}$, we conclude that the singularities of $y_0$ in ${\cal A}_{j,0}$ in sheet $1$ and the singularities of $y_{m+1}$ in ${\cal A}_{j,m+1}$ in sheet $m+1$ can be only poles, not cuts.

Since condition \ref{genus_ss} requires that the genus be $\mathfrak g=nm$, we see that there can be no other cuts than the ones already discussed -- the genus would be higher, otherwise.

\subsection{Symplectic change of functions}

The spectral curve of the matrix model is $\spcurve_{MM} = (\curve,x,y_0)$, and our goal is to relate it to the mirror curve described in section \ref{secmirrorcurve}. The mirror curve is described via the algebraic equation (\ref{mirror_curve}) in the two functions $x_1,x_2: \curve_{mirror} \rightarrow \IC \IP^1$  (in the patch $x_0=1$). We wish to obtain a similar algebraic description of $\curve$. Due to log singularities in $y_0$, to be traced to the small $g_s$ behavior of $\psi_q(x)$, an algebraic equation in the variables $(x,y_0)$ cannot exist (recall that $x$ is meromorphic). In this section, we shall, via a series of symplectic transformations on the $y_i$ of the type enumerated in section \ref{secdefspinvs}, arrive at functions $Y_i$ that are meromorphic on $\curve$, and hence each present a viable candidate to pair with $x$ to yield an algebraic equation for $\curve$.

\bigskip

Essentially, we wish to introduce the exponentials of $y_i$. While this will eliminate the log singularities, poles in $y_i$ would be elevated to essential singularities. We hence first turn to the question of eliminating these poles.

\subsubsection{The arctic circle property} \label{arctic}
On the physical sheet, the interpretation of a pole of $y_i$ is as an eigenvalue of the matrix $M_i$ with delta function support. Such a so-called frozen eigenvalue can arise in the following way:

The sum over all partitions is dominated by partitions close to a typical equilibrium partition, i.e. a saddle point. The typical partition has a certain typical length referred to as its equilibrium length $\bar{n}$. All partitions with a length very different from the equilibrium length contribute only in an exponentially small way (and thus non-perturbatively) to the full partition function. Introducing a cutoff on the length of partitions which is larger than the equilibrium length hence does not change the perturbative part of the partition function. Now recall that when we defined the $h_i(\gamma)$ of a representation $\gamma$ in appendix \ref{our_matrix_model}, we introduced an arbitrary maximal length $d$ such that $l(\gamma)\leq d$ and set
\beq
h_i(\gamma) = a_\gamma + d -i + \gamma_i  \,.
\eeq
Setting $\gamma_i =0$ for $d \ge i > \bar{n}$ yields $h_i$ that do not depend on the integration variables, hence are frozen at fixed values. This behavior is referred to as the arctic circle property \cite{Johansson}, as all eigenvalues beyond the arctic circle situated at equilibrium length $\bar{n}$ are frozen.

Returning to our matrix model, the eigenvalues of the matrices $M_i$ are given by $q^{(h_{j,i})_l}$. For $d \ge l > n_{j,i}$, they are frozen, and thus contribute poles to $y_i$ by (\ref{resolvent})  (recall that poles of the resolvent correspond to eigenvalues with delta function support) in the physical sheet. We will assume that these are the only poles in the physical sheet and we subtract them to obtain new functions $\td y_i$,
\ba
\td y_0(z) =x(z)y_0(z)- \sum_j \sum_{k=0}^{d-1} {g_sx(z)\over x(z)-q^{a_{j,0}+k}} \,,
\ea
\ba
\td y_{m+2}(z) = x(z)y_{m+2}(z)+ \sum_j \sum_{0}^{d-1} {g_s x(z)\over x(z)-q^{a_{j,m+1}+k}}
\ea
and for $i=1,\dots,m+1$,
\ba
\td y_i(z)=x(z)y_i(z) -\sum_j \sum_{k=0}^{d-n_{j,i}-1} {g_sx(z)\over x(z)-q^{a_{j,i}+k}} + \sum_j \sum_{k=0}^{d-n_{j,i-1}-1} {g_s x(z)\over x(z)-q^{a_{j,i-1}+k}}  \,.
\ea
We have set
\beq
n_{j,0}=0
\,\, , \quad
n_{j,m+1}=0 \,.
\eeq
Notice that at large $x(z)$ in sheet $i$ we have
\beq
\td y_0 \sim  O(1/x(z))
\,\,\, , \qquad
\td y_{m+2} \sim  O(1/x(z))
\eeq
and for $i=1,\dots,m+1$
\beq
\td y_i \sim   g_s \sum_j (n_{j,i}-n_{j,i-1}) + O(1/x(z)).
\eeq

\medskip

As a general property of $\psi_q$, we have for any integer $n_{j,i}\leq d$ 
\beq
\psi_q(q^{a_{j,i}}/x) = \psi_q(q^{a_{j,i}+d-n_{j,i}}/x)+ \sum_{k=0}^{d-n_{j,i}-1} {x\over x-q^{a_{j,i}+k}}   \,.
\eeq
Hence, the loop equations for the new functions $\tilde{y}_i$ read
\ba
\td y_{i+1}(z)-\td y_{i}(z) 
&=& i\pi\, + g_s \, S_i  \cr
&& + g_s \sum_j (2\psi_q(q^{a_{j,i}+d-n_{j,i}}/x(z)) - \psi_q(q^{a_{j,i+1}+d-n_{j,i+1}}/x(z)) \cr
&& -\psi_q(q^{a_{j,i-1}+d-n_{j,i-1}}/x(z)))  + g_s\, {x(z) f_i'(x(z))\over f_i(x(z))}   \,,\cr
\td y_1(z)-\td y_{0}(z) 
&=& g_s \sum_j \psi_q(q^{a_{j,0}+d}/x(z))-g_s\sum_j\psi_q(q^{a_{j,1}+d-n_{j,1}}/x(z)) \,,
\cr
\td y_{m+2}(z) -\td y_{m+1}(z)
&=&  g_s \sum_j \psi_q(q^{a_{j,m+1}+d}/x(z))-g_s\sum_j\psi_q(q^{a_{j,m}+d-n_{j,m}}/x(z)) \,.
\ea

\medskip

\begin{figure}[t]
 \centering
  \includegraphics[width=14cm]{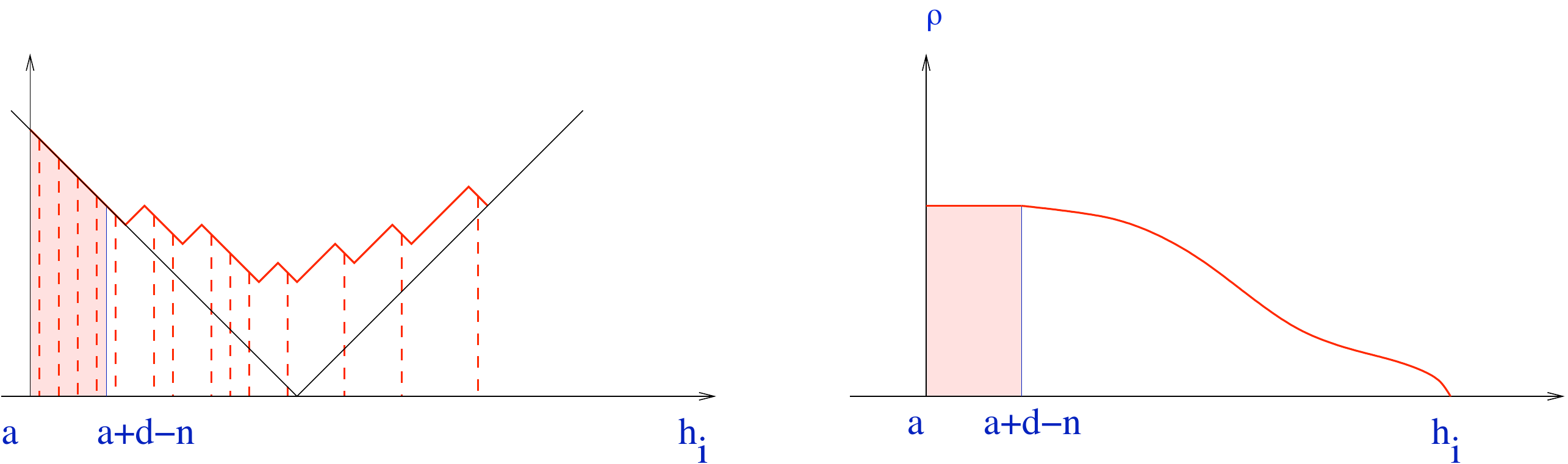}
 \caption{\footnotesize{We shift the cut-off $d$ on the representation lengths, $d\to n_{j,i}$, with $n_{j,i}$ chosen such that frozen eigenvalues in the expected distribution of the $h_i$ are suppressed. In the limit of vanishing spacing ($g_s\to 0$), the equidistant frozen eigenvalues give rise to a constant eigenvalue density region.}}
 \label{figctedensityshiftnij}
\end{figure}

The $n_{j,i}$ in the above definitions are defined as the equilibrium lengths, i.e. by the property that the functions $\tilde{y}_i$ have no poles on their physical sheet. That such a choice of $n_{j,i}$ exists is suggested by the arctic circle property.

Note that the $n_{j,i}$ can also be specified by the fact that $q^{a_{j,i} + d - n_{j,i}}$ be the beginning of the cut encircled by $\gamma_{j,i}$. As we have identified the discontinuities of $y_i$ to lie across branchcuts of $x$, this implies that $x$ has ramification points at the element of $x^{-1}(q^{a_{j,i} + d - n_{j,i}})$ lying on the physical sheets of $y_i$.

Note that the arctic circle property also implies the perturbative independence of our expressions from the arbitrary cut-off $d$. Changing $d$ to $d+d'$ merely introduces $d'$ new frozen eigenvalues $h_i$. This independence from $d$ is important in establishing the equality between the topological string partition function and our matrix integral (\ref{m_integral}), as the topological vertex formulae in fact are formulated in the limit $d\rightarrow \infty$.

\subsubsection{Obtaining globally meromorphic functions}

We have arrived at functions $\tilde{y_i}$ that have no poles on their physical sheet, and are thus safely exponentiated there. We wish now to use the loop equations to obtain functions which are globally well-behaved. 

To this end, we note that since the Gromov-Witten invariants are defined as a formal power series in $g_s$, we can compute the spectral curve order by order in $g_s$, invoking the following small $\ln q$ expansion \cite{part_1}:
\ba
\psi_q(q^{a_{j,i}+d-n_{j,i}}/x)
&\sim& -\frac{1}{g_s} \ln(1-\frac{x}{q^{a_{j,i}+d-n_{j,i}}}) +\frac{x}{2(x-q^{a_{j,i}+d-n_{j,i}})} \cr
&& +\frac{1}{g_s}\sum_{n=1}^\infty {B_{2n}\,g_s^{2n}\over (2n)!}\,\Li_{1-2n}(q^{d-n_{j,i}+a_{j,i}}/x)  \,.
\ea
The functions $f'_i/f_i$ are completely non-perturbative; one can
easily check with the above expansion that they can be replaced by $0$ to every order in $g_s$.

Introducing new functions $X(z)$ and $Y_i$ by the formulae
\beq
x(z) = q^d\, X(z) \,,
\eeq
\beq
\td y_0(z) = \ln{Y_0(z)} \,,
\eeq
\beq
\td y_{m+2}(z) = \ln{Y_{m+2}(z)} \,,
\eeq
and for $i=1,\dots,m+1$
\ba
\td y_i(z) 
&=& \ln{Y_i(z)} + \sum_j{X(z) g_s\over 2(X(z)-q^{a_{j,i}-n_{j,i}})} +\frac{1}{g_s} \sum_j \sum_{n=1}^\infty {B_{2n}\, g_s^{2n}\over (2n)!}\, \Li_{1-2n}(q^{a_{j,i}-n_{j,i}}/X(z)) \cr
&& -  \sum_j{X(z) g_s\over 2(X(z)-q^{a_{j,i-1}-n_{j,i-1}})} - \frac{1}{g_s}\sum_j \sum_{n=1}^\infty {B_{2n}\, g_s^{2n}\over (2n)!}\, \Li_{1-2n}(q^{a_{j,i-1}-n_{j,i-1}}/X(z))
\ea

yields loop equations that are algebraic on their right hand side,

\ban \label{ratios} 
\frac{Y_i}{Y_{i+1}}
&=& -\,
e^{-g_s \,S_i}\prod_j \frac {(X-q^{a_{j,i+1}-n_{j,i+1}})(X-q^{a_{j,i-1}-n_{j,i-1}})}{(X-q^{a_{j,i}-n_{j,i}})^2} \cr
&& \qquad \prod_j q^{2(a_{j,i}-n_{j,i})-(a_{j,i+1}-n_{j,i+1})-(a_{j,i-1}-n_{j,i-1})} \,,\cr
\frac{Y_0}{Y_1} &=&\prod_j \frac {(X-q^{a_{j,1}-n_{j,1}})}{(X-q^{a_{j,0}})} \prod_j q^{a_{j,0}-(a_{j,1}-n_{j,1})} \,,\cr
\frac{Y_{m+1}}{Y_{m+2}} &=&\prod_j \frac {(X-q^{a_{j,m}-n_{j,m}})}{(X-q^{a_{j,m+1}})} \prod_j q^{a_{j,m+1}-(a_{j,m}-n_{j,m})} \,,
\ean
i.e.
\ba
{Y_i\over Y_0} 
= e^{g_s(S_1+\dots+S_{i-1})}\,\prod_j q^{(a_{j,i}-n_{j,i})-(a_{j,i-1}-n_{j,i-1})}\, \prod_j {X-q^{a_{j,i-1}-n_{j,i-1}}\over X-q^{a_{j,i}-n_{j,i}}}  \,.
\ea
Since we have argued that the $Y_i$ are holomorphic on their physical sheet, and the ratio $Y_i/Y_{i+1}$ is purely algebraic, we conclude that the $Y_i$ are meromorphic functions on all of $\curve$. This was the goal we had set out to achieve.

Note that the above changes of variables have modified the asymptotics at infinity and the integrals over the $\cal{A}$-cycles. More precisely, we have
\beqn \label{asymptotics} 
\forall i \in [1,m+1]: \, \ln Y_i \mathop{\sim}_{\infty_i}  \td y_i \mathop{\sim}_{\infty_i} g_s\sum_j (n_{j,i}-n_{j,i-1}) +O(\frac{1}{x})\,,
\eeqn
\beq \ln Y_0 = \td y_0 \mathop{\sim}_{\infty} O(\frac{1}{x})\,,
\eeq
\beq \ln Y_{m+2} = \td y_{m+2} \mathop{\sim}_{\infty} O(\frac{1}{x}) \,.
\eeq
The filling fraction equation reads
\beqn \label{integrals} \frac{1}{2i\pi} \int_{{\cal{A}}_{j,i}} \frac{\td y_i(z)}{x(z)}dx(z)=g_s(d-(d-n_{j,i}))= g_s\,n_{j,i} \,.
\eeqn
In terms of $Y_0$, these conditions can be rewritten as
\beq
\ln Y_0 \mathop{\sim}_{\infty_i}   -g_s(S_1+\dots+S_{i-1}) + g_s\sum_{j=0}^n (a_{j,i}-a_{j,i-1}) +O(\frac{1}{x})
\eeq
and
\ba
\frac{1}{2i\pi} \int_{{\cal{A}}_{j,i}} \ln {Y_0}\, \frac {dX}{X}
&=& \frac{1}{2i\pi} \int_{{\cal{A}}_{j,i}} \ln {Y_i}\, \frac {dX}{X} + \frac{1}{2i\pi} \int_{{\cal{A}}_{j,i}} \ln {X}\,\, d \ln \left(\frac{Y_i}{Y_0} \right) \cr
&=& \frac{1}{2i\pi} \int_{{\cal{A}}_{j,i}} \ln {Y_i}\, \frac {dX}{X} + \frac{1}{2i\pi} \int_{{\cal{A}}_{j,i}} \ln {X}\,\, d \left( \sum_{k=1}^i \ln \frac{Y_k}{Y_{k-1}} \right) \cr
&=& \frac{1}{2i\pi} \int_{{\cal{A}}_{j,i}} \ln {Y_i}\, \frac {dX}{X} - \frac{1}{2i\pi} \int_{{\cal{A}}_{j,i}} \ln {X}\,\,\left( \sum_l {dX\over X-q^{a_{l,i}-n_{l,i}}} - {dX\over X-q^{a_{l,i-1}-n_{l,i-1}}}\right) \cr
&=& \frac{1}{2i\pi} \int_{{\cal{A}}_{j,i}} \ln {Y_i}\, \frac {dX}{X} + g_s(a_{j,i}-n_{j,i}) \cr
&=& g_s n_{j,i}+ g_s\,(a_{j,i}-n_{j,i}) \cr
&=& g_s a_{j,i}  \,.
\ea

\subsection{Recovering the mirror curve}

We have argued above that $X$ and $Y_i$, and hence in particular
$Y_0$, are meromorphic functions on $\curve$. There must hence exist a
polynomial $H(X,Y)$ such that (see e.g. Theorem 5.8.1 in \cite{Jost})
\beq
H(X,Y_0)=0  \,.
\eeq
The facts that $X$ provides an $m+1$ sheeted cover of $\IC \IP^1$ and that $Y_0$ may have $n+1$ poles in its physical sheet imply that the polynomial $H$ has degrees at least $(n+1,m+1)$. As above, we shall choose the minimal hypothesis that it has exactly these degrees. Thus,
\beqn  \label{mirror_curve_H}
H(X,Y) = \sum_{i=0}^{m+1}\sum_{j=0}^{n+1} H_{i,j} X^j \, Y^i.
\eeqn
As we saw in section \ref{secmirrorcurve}, projectivizing a generic polynomial of these degrees (yielding a homogeneous polynomial of degree $m+n+2$) indeed gives rise to a curve of genus ${\mathfrak g}=nm$.

We now need to determine the $(n+2)(m+2)-1$ unknown coefficients of $H$ ($H$ is defined up to a global multiplicative constant).

\bigskip

The cycle integrals 
\beq
\oint_{{\cal A}_{j,i}} \ln{Y_0}\, \, {dX\over X} = 2i\pi g_s\,\, a_{j,i}
\eeq
provide $(n+1)m$ constraints on the coefficients of $H$.
We also have $m+1$ constraints for the behavior at $\infty_i$, $i=1, \ldots, m+1$,
\beq
\Res_{\infty_i} \ln{Y_0}\, \, {dX\over X} = g_s(S_1+\dots+S_{i-1}) - g_s\sum_{j=0}^n (a_{j,i}-a_{j,i-1})  \,.
\eeq
Finally, requiring that $Y_0$ has poles at $q^{a_{j,0}}$ and $Y_{m+2}$ has zeroes at $q^{a_{j,m+1}}$ gives another $2(n+1)$ constraints, which we may write as
\beq
\Res_{q^{a_{j,0}}} \ln{X}\, \, {dY_0\over Y_0} = g_s a_{j,0} \,,
\eeq
\beq
\Res_{q^{a_{j,m+1}}} \ln{X}\, \, {dY_{m+2}\over Y_{m+2}} = g_s a_{j,m+1}  \,.
\eeq
This gives enough equations to completely determine $H$.
Knowing $H$, we know the location of branch points as functions of $a_{j,i}$'s and $S_i$'s, and can hence determine the $n_{j,i}$ by requiring that $q^{a_{j,i}-n_{j,i}}$ be a branch point.

\medskip
Notice that we can choose to express the period integrals in any linear combination of ${\cal A}$-cycles. In particular,
\beq
\oint_{{\cal A}_{j,i+1}-{\cal A}_{j,i}} \ln{Y_0}\, \, {dX\over X} = 2i\pi g_s\,\, (a_{j,i+1}-a_{j,i}) = 2i\pi\,\, g_s\, t_{j,i}  \,,
\eeq
\beq
\oint_{{\cal A}_{j,i+1}-{\cal A}_{j+1,i}} \ln{Y_0}\, \, {dX\over X} = 2i\pi g_s\,\, (a_{j,i+1}-a_{j+1,i}) = 2i\pi\,\, g_s\, r_{j,i}  \,.
\eeq
Similarly, we may also take linear combinations of ${\cal A}$-cycles together with circles surrounding the poles or zeroes of $x$ in order to get the $s_{j,i}$ classes. We hence conclude that the periods of the curve $H(X,Y_0)=0$ yield the quantum corrected K\"ahler parameters of the fiducial toric geometry $\CYX_0$, allowing us to identify it with the corresponding mirror curve.

\begin{figure}[h]
 \centering
  \includegraphics[width=5cm]{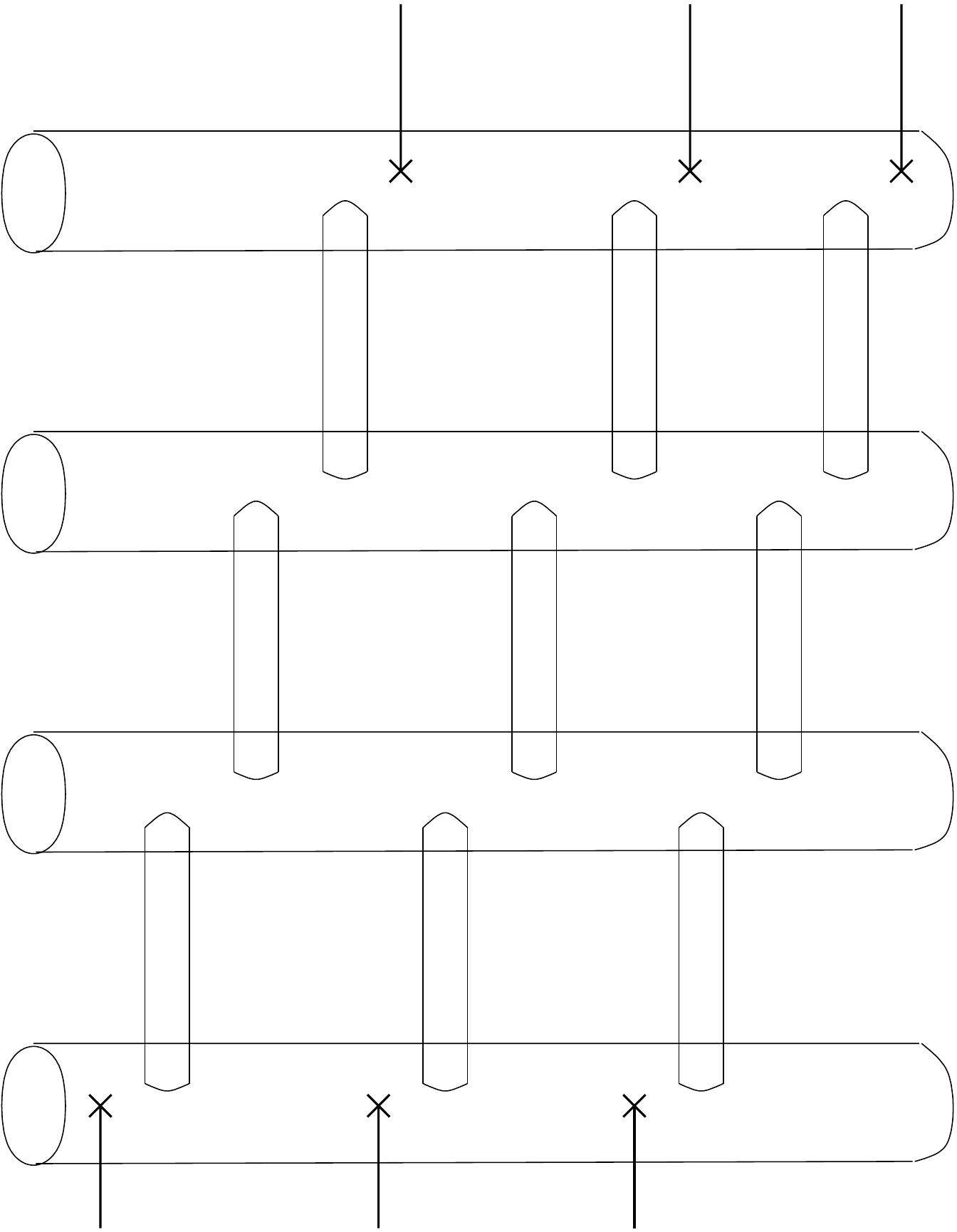}
 \caption{\footnotesize{The spectral curve $(X,{1\over X}\ln{(Y_0)})$ has the following structure:
 $X(z)$ is a meromorphic function of degree $m+1$ on a curve of genus ${\frak{g}}=nm$. Therefore it has $m+1$ poles and $m+1$ zeroes. It provides a branched covering of $\IC \IP^1$. We prefer to represent $\ln X$ instead of $X$, and thus we have $m+1$ copies of the $\ln X$-cylinder. In each sheet there is one zero and one pole of $X$. $Y_0$ is a meromorphic function of degree $n+1$, so that it has $n+1$ zeroes in sheet $0$, and $n+1$ poles in sheet $m+1$.
We recognize the mirror curve ${\cal S}_{\hat\CYX_0}$,  which is a thickening of the toric web diagram.
 }}
 \label{figspcurveMM1}
\end{figure}

\subsection{Topological expansion and symplectic invariants}

Following \cite{EPrats}, we obtained
\beq
\spcurve_{MM}=(\curve,x,y_0) 
\eeq
as the spectral curve of our matrix model at the end of section \ref{applying_the_rules}.

As reviewed in section \ref{secdefspinv}, we can compute the corresponding symplectic invariants $F_g(\check\spcurve_{MM})$, which assemble to yield the matrix model partition function \cite{EPrats},
\beq
\ln Z = \sum_g g_s^{2g-2}\, F_g(\spcurve_{MM}) \,.
\eeq
The symplectic transformation $\mathfrak{R}$ of section \ref{secdefspinvs} maps $(\curve, x, y_0)$ to $(\curve,x,\frac{1}{x}\ln Y_0)$ order by order in $g_s$. $\mathfrak{F}$ maps this to $(\curve,X,{1\over X}\ln Y_0)$, and a second application of $\mathfrak{F}$ yields
\beq
\hat\spcurve_{MM}=(\curve,\ln{X},\ln Y_0) \,.
\eeq
By the symplectic invariance of the $F_g$, we therefore have, order by order in powers of $g_s$,
\beq
F_g(\spcurve_{MM}) = F_g(\hat\spcurve_{MM}) \,.
\eeq
Since our matrix model was engineered to reproduce the Gromov-Witten invariants of $\CYX_0$, we have arrived at
\beq
GW_g(\CYX_0) = F_g(\curve,\ln{X},\ln Y_0)  \,,
\eeq
with $X$ and $Y_0$ obeying the algebraic equation
\beq
H_0(X,Y_0)=0
\eeq
which coincides with the equation (\ref{mirror_curve}) describing the mirror curve of $\CYX_0$.

Given our minimality assumptions on the spectral curve, we have thus derived the BKMP conjecture for the fiducial geometry $\CYX_0$.

\subsection{The small $q$ limit and the thickening prescription}
\label{secproofsp}

The above derivation of the spectral curve for the matrix model is not fully rigorous, as we have relied on making minimal assumptions along the way. Although the spectral curve we have found here satisfies all the constraints of section \ref{secspcurvegenchain}, to prove that it is the spectral curve of our matrix model requires a uniqueness result which we currently do not have.

In this section, we provide a heuristic argument that the qualitative behavior of the spectral curve and the mirror curve coincide at small $q$.

\begin{figure}[h]
 \centering
  \includegraphics[width=7cm]{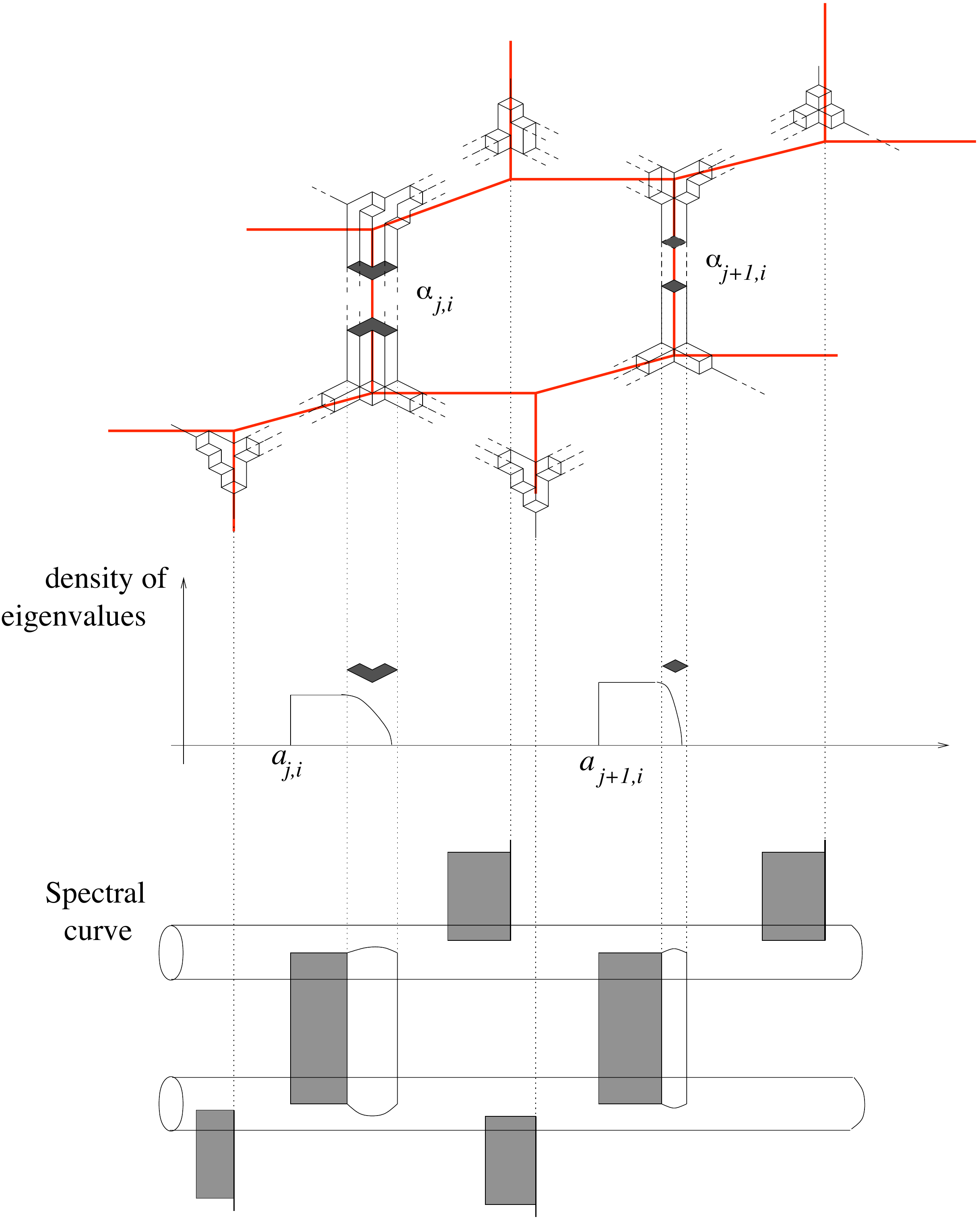}
 \caption{\footnotesize{ In the small $q$ limit, only very small partitions contribute to the matrix integral, therefore the density of eigenvalues of $M_i$ tends to the flat density (a Dirac comb of equidistant delta functions), the non-flat part, which reflects the cuts of the spectral cut, shrinks to zero.}}
 \label{figsmallqpplane}
\end{figure}

At small $q$, only very small partitions contribute to the matrix integral. Almost all eigenvalues of $M_i$ are frozen to the values $q^{a_{j,i}+d-l}$. By the arguments in section \ref{arctic}, the resolvent $W_i(x)$ hence behaves at small $q$ as
\beq
W_i(x) \sim \sum_{j=0}^n \sum_{l=1}^d\,\, {1\over x-q^{a_{j,i}-l+d}} + {\rm small\, cut\, near\,} q^{a_{j,i}+d}  \,.
\eeq
Pictorially, the size of the cuts is shrinking in this limit, replacing the spectral curve by its skeleton, see figure \ref{figsmallqpplane}.

On the other hand, the mirror curve is a priori a tree level quantity, hence does not depend on $q=e^{-g_s}$. However, recall that we have defined the K\"ahler parameters $Q$ associated to a curve $\curve$ as 
\beq
Q= q^{\int_{\curve} J} \,.
\eeq
The large $q$ limit hence corresponds to the large curve class limit, i.e. the distance between the vertices of the pairs of pants out of which the mirror curve is constructed is taken to infinity. Just as the spectral curve, the mirror curve thus collapses to its skeleton in the $q \rightarrow 0$ limit.
\section{The general BKMP conjecture}  \label{finishing_proof}

So far, we have obtained the BKMP conjecture only for the fiducial geometry $\CYX_0$. Studying the behavior of the partition function under flop transitions will allow us to extend our argument to arbitrary toric geometries.

\subsection{Flop invariance of toric Gromov-Witten invariants}

\begin{figure}[h]
 \centering
 \includegraphics[width=6cm]{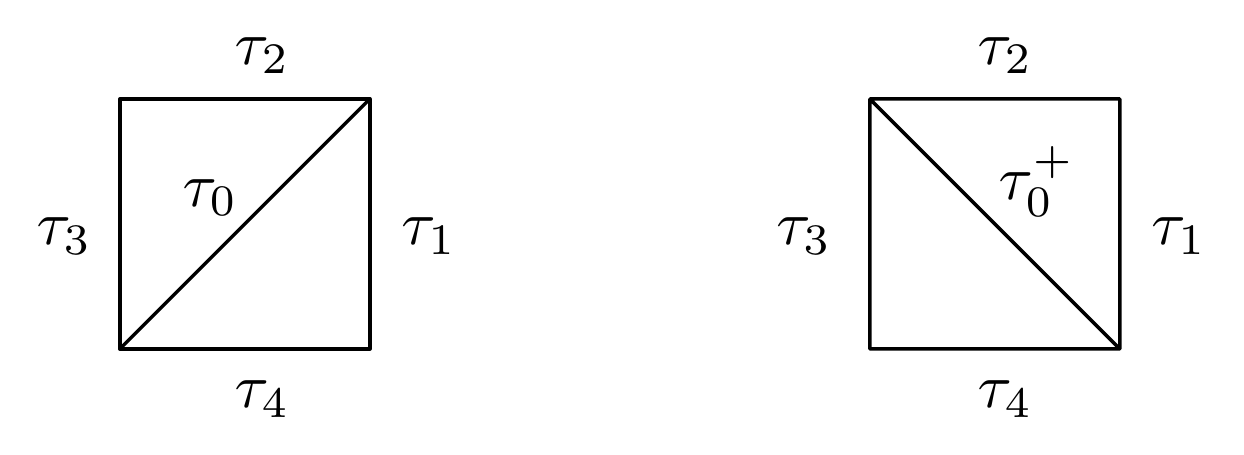}
 \caption{\footnotesize{$\CYX$ and $\CYX^+$ in the vicinity of the (-1,-1) curve.}}
 \label{toric_flopped}
\end{figure}

Under the proper identification of curve classes, Gromov-Witten invariants (at least on toric manifolds) are invariant under flops. Assume the toric Calabi-Yau manifolds $\CYX$ and $\CYX^+$ are related via a flop transition, $\phi: \CYX \rightarrow \CYX^+$. In a neighborhood of the flopped $(-1,-1)$ curve, the respective toric diagrams are depicted in figure \ref{toric_flopped}.

The 1-cones of $\Sigma_\CYX$, corresponding to the toric invariant divisors of $\CYX$, are not affected by the flop, hence can be canonically identified with those of $\CYX^+$. The 2-cones $\tau_i$ in these diagrams correspond to toric invariant 2-cycles $C_i$, $C_i^+$ in the geometry. The curve classes of $\CYX$ push forward to classes in $\CYX^+$ via
\ban\label{flopping}
\phi_*([C_0]) = - [C_0^+] \,, \quad \phi_*([C_i]) =  [C_i^+] + [C_0^+] \,. 
\ean
All other curve classes $\vec{C}$ of $\CYX$ are mapped to their canonical counterparts in $\CYX^+$. Under appropriate analytic continuation and up to a phase factor (hence the $\propto$ in the following formula), the following identity then holds \cite{Witten_Phases,IqbalKashaniPoor, KonishiMinabe},
\ba
Z_{GW}(\CYX,Q_0,Q_1,\ldots,Q_4,\vec{Q}) \propto Z_{GW}(\CYX^+,1/Q_0, Q_0 Q_1, \ldots, Q_0 Q_4,\vec{Q}) \,,
\ea
i.e.
\ba
GW_g(\CYX,Q_0,Q_1,\ldots,Q_4,\vec{Q}) = GW_g(\CYX^+,1/Q_0, Q_0 Q_1, \ldots, Q_0 Q_4,\vec{Q}) \,.
\ea

\subsection{Proof of flop invariance via mirror symmetry}
Flop invariance of Gromov-Witten invariants upon the identification (\ref{flopping}) is immediate upon invoking mirror symmetry, as (\ref{flopping}) maps the mirror curve of $\CYX$ to that of $\CYX^+$. The proof is a simple computation.

\begin{figure}[h]
 \centering
 \includegraphics[width=6cm]{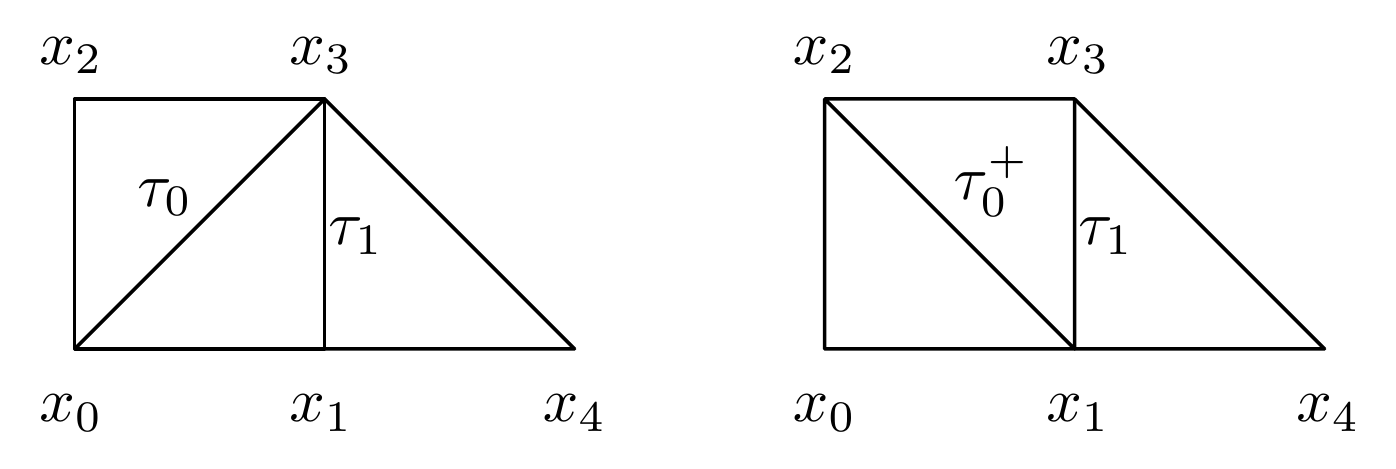}
 \caption{\footnotesize{$\CYX$ and $\CYX^+$ in the vicinity of the (-1,-1) curve.}}
 \label{mirror_flopped}
\end{figure}
Let us introduce the notation $t_0, t_1, t_0^+, t_1^+$ for the K\"ahler volume of the curve classes $C_i,C_i^+$ corresponding to the respective 2-cones.
In terms of these, we obtain for the mirror curve of $\CYX$
\beq
x_0 + x_1 + x_2 + \frac{x_1 x_2}{x_0} e^{T_0} + \frac{x_1^2}{x_0} e^{-T_1} = 0 \,,
\eeq
while the mirror curve of $\CYX^+$ is given by
\beq
x_0 + x_1 + x_2 + \frac{x_1 x_2}{x_0} e^{-T_0^+} + \frac{x_1 x_3}{x_2} e^{-T_1^+} = 0\,.
\eeq
Upon invoking $x_3 = \frac{x_1 x_2}{x_0}e^{-T_0^+}$, we easily verify that the identification (\ref{flopping}) maps these curves and their associated meromorphic 1-forms $\lambda$ into each other. 

\subsection{The BKMP conjecture}
\begin{figure}[h]
 \centering
 \includegraphics[width=10cm]{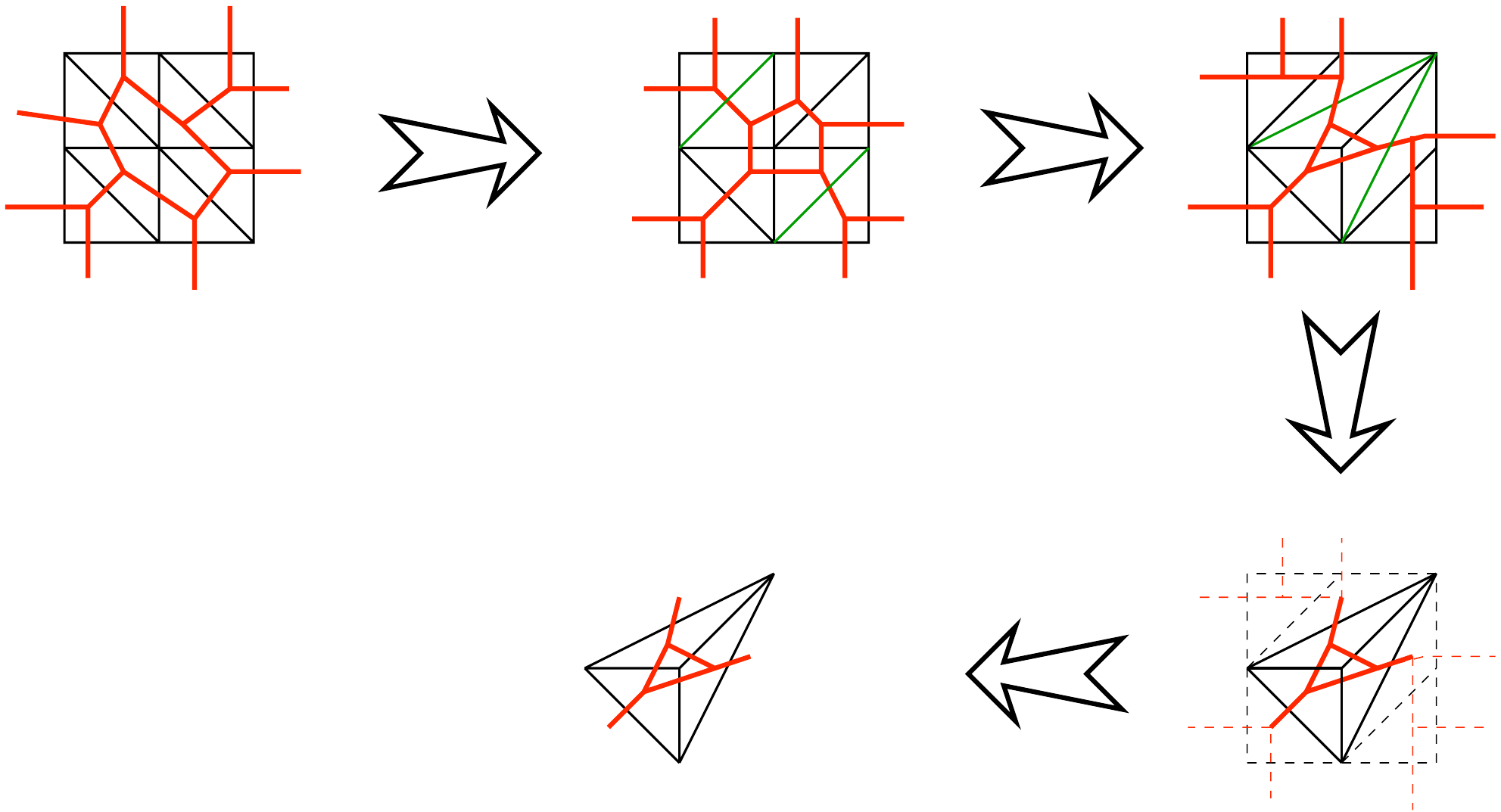}
 \caption{\footnotesize{Example: We obtain local $\mathbb P^2$ from the fiducial geometry with $2\times 2$ boxes by performing five flops and then sending the K\"ahler parameters of the unwanted edges to $\infty$.}}
 \label{flopP2}
\end{figure}

Any toric Calabi-Yau manifold $\CYX$ with K\"ahler moduli $\vec{Q}$ can be obtained from a sufficiently large fiducial geometry $(\CYX_0,\vec{Q}_0)$ upon performing a series of flop transitions and taking unwanted K\"ahler moduli of $\CYX_0$ to $\infty$, see figure \ref{flopP2} for an example.

The K\"ahler moduli of $\CYX$ are related to those of $\CYX_0$ by some relation $\vec{Q}=f(\vec{Q}_0)$. We have just argued that the mirror curves of $\CYX_0$ and $\CYX$ are equal upon this identification,
\beq
\spcurve_{\CYX,\vec{Q}} = \spcurve_{\CYX_0,\vec{Q_0}} \,,
\eeq
as are the respective Gromov-Witten invariants,
\beq
GW_g(\CYX,\vec{Q}) = GW_g(\CYX_0,\vec{Q}_0) \,.
\eeq
Given the BKMP conjecture for the fiducial geometry,
\beq
GW_g(\CYX_0,\vec{Q}_0)
=F_g(\spcurve_{\CYX_0,\vec{Q}_0}) \,,
\eeq
its validity thus follows for any toric Calabi-Yau manifold:
\beq
\encadremath{
GW_g(\CYX,\vec{Q})
=F_g(\spcurve_{\CYX,\vec{Q}})
}\eeq

\section{Conclusion}
Taking our matrix model from \cite{part_1} as a starting point and imposing certain minimality conditions on the spectral curve, we have thus derived the BKMP conjecture, for closed topological strings, for all toric Calabi-Yau manifolds in the large radius limit. As we have emphasized throughout, elevating our procedure to a formal proof of the conjecture requires a more rigorous derivation of the spectral curve of our matrix model.

It should also be possible to extend our argument to
open Gromov-Witten invariants by invoking loop operators, which relate
closed to open invariants. In \cite{EOFg}, such an operator was defined in the matrix model context. An analogous operator should also exist in the theory of Gromov-Witten invariants \cite{Cavalieri}. Establishing the equivalence of these two loop operators would allow us to conclude that the $W_n^{(g)}$'s of the spectral curve $\spcurve_{\CYX}$ are the open Gromov-Witten invariants of $\CYX$.

Finally, our treatment of the BKMP conjecture took place at large radius. One should study the behavior of the matrix model as one moves away from large radius e.g. to orbifold points, and see whether the phase transitions of the topological string are captured accurately by the matrix model. Of course, the main tool on the topological string side employed in this work, the topological vertex, is no longer applicable in these regions of moduli space.

\section*{Acknowledgments}
B.E. and O.M. would like to thank M. Bertola, J. Harnad, V. Bouchard,  M. Mari\~ no, M. Mulase,  H.~ Ooguri, N.~Orantin, B. Safnuk, for useful and fruitful discussions on this subject. A.K. would like to thank Vincent Bouchard and Ilarion Melnikov for helpful conversations. The work of B.E. is partly supported by the Enigma European network MRT-CT-2004-5652, ANR project GranMa "Grandes Matrices Al\'eatoires" ANR-08-BLAN-0311-01,  
by the European Science Foundation through the Misgam program,
by the Quebec government with the FQRNT. 
B.E. would like to thank the AIM, as well as the organizers and all participants to the workshop \cite{AIM} held at the AIM june 2009.
O.M. would like to thank the CRM (Centre de recheche math\'ematiques de Montr\'eal, QC, Canada) for its hospitality.

\appendix{}

\section{The matrix model}   \label{our_matrix_model} 
In this appendix, which is mainly a reprint of section 4 of \cite{part_1}, we present the matrix model which reproduces the topological string partition function on the fiducial geometry $\CYX_0$, and whose spectral curve we derive in the text.

SConsider the fiducial geometry $\CYX_0$ of size $(n+1)\times (m+1)$, with K\"ahler parameters $t_{i,j}=a_{i,j}-a_{i,j+1}$, $r_{i,j} = a_{i,j+1}-a_{i+1,j}$, and  $s_{i,j}$, as depicted in figure \ref{fiducial_geometry}. We write
\beq
\vec a_i  = (a_{0,i},a_{1,i},\dots,a_{n,i}).
\eeq

Assume that the external representations are fixed to $\vec\alpha_{m+1} = (\alpha_{0,m+1},\alpha_{1,m+1},\dots,\alpha_{n,m+1})$ on the upper line, and $\vec\alpha_0 = (\alpha_{0,0},\alpha_{1,0},\dots,\alpha_{n,0})$ on the lower line. For the most part, we will choose these to be trivial.

\bigskip

We now define the following matrix integral ${\cal Z}_{\rm MM}$ (${}_{\rm MM}$ for Matrix Model),
\ban
{\cal Z}_{\rm MM}(Q,g_s,\vec\alpha_{m+1},\vec\alpha_0^T)
&=& \Delta(X(\vec \alpha_{m+1}))\,\, \Delta(X(\vec \alpha_0)) \,\, 
\prod_{i=0}^{m+1} \int_{H_N(\Gamma_i)} dM_i \,
 \prod_{i=1}^{m+1}\int_{H_N({\mathbb R}_+)}\,dR_i \nn \\
&& \prod_{i=1}^{m} e^{{-1\over g_s}\,\tr \left[ V_{\vec a_i}(M_i)-V_{\vec a_{i-1}}(M_i) \right]
} \,\,\,
 \prod_{i=1}^{m} e^{{-1\over g_s}\,\tr \left[V_{\vec a_{i-1}}(M_{i-1})-V_{\vec a_{i}}(M_{i-1}) \right]
} \nn \\
&& \prod_{i=1}^{m+1} e^{{1\over g_s} \tr (M_i-M_{i-1})R_i} \,\,\,
 \prod_{i=1}^{m} e^{(S_i+{i\pi\over g_s})\,\tr\, \ln M_i}\,  \nn\\
&& e^{\tr \ln f_{0}(M_0)}\,\,e^{\tr \ln f_{m+1}(M_{m+1})}\,\, \prod_{i=1}^{m} e^{\tr \ln f_{i}(M_i)} \,. 
\ean
All matrices are taken of size
\beq
N=(n+1)\, d \,.
\eeq
$d$ denotes a cut-off on the size of the matrices, on which, as discussed in section \ref{arctic}, the partition function depends only non-perturbatively.
We have introduced the notation
\beq
X(\vec \alpha_{m+1})  = {\rm diag} (X(\vec \alpha_{m+1})_i)_{i=1,\dots,N}
\,\, , \qquad
X(\vec \alpha_{m+1})_{j d+k} = q^{h_k(\alpha_{j,m+1})},
\eeq
\beq
X(\vec \alpha_0)  = {\rm diag} (X(\vec \alpha_0)_i)_{i=1,\dots,N}
\,\, , \qquad
X(\vec \alpha_0)_{j d+k} = q^{h_k(\alpha_{j,0})},
\eeq
for $k=1, \ldots, d$, $j=0, \ldots,n$, where
\beqn \label{defh}
h_i(\gamma)=\gamma_i-i+d+a \,.
\eeqn
$\Delta(X)=\prod_{i<j}(X_i-X_j)$ is the Vandermonde determinant. The potentials $V_{\vec a_i}(x)$ are given by
\beqn    \label{partpot}
V_{\vec a}(X) = -g_s\,\sum_{j=0}^n \ln{\left(g(q^{a_j}/X)\right)} 
\eeqn
in terms of the $q$-product
\ba
g(x) = \prod_{n=1}^\infty (1-{1\over x}\,q^n) \,.
\ea
For $i=1,\dots,m$, we have defined
\beq
f_i(x) =  \prod_{j=0}^n {g(1)^2\,\,e^{({1\over 2}+{i\pi\over \ln q})\, \ln{(x q^{1-a_{j,i}})}}\, \,e^{{(\ln{(x q^{1-a_{j,i}})})^2\over  2 g_s}}\over g(x\,q^{1-a_{j,i}})\, g(q^{a_{j,i}}/x)\,} \,.
\eeq
The denominator of these functions induces simple poles at $x=q^{a_{j,i}+l}$ for $j=0,\dots,n$ and $l\in \mathbb Z$. The numerator is chosen such that they satisfy the relation $f_i(qx)=f_i(x)$. This enforces a simple $l$-dependence of the residues taken at $x=q^{a_{j,i}+l}$, given by a prefactor $q^l$ -- a fact which will be important in the following. These residues are in fact given by
\beqn \label{resf}
 \Res_{q^{a_{j,i}+l}} f_i(x) = q^{a_{j,i}+l}\,\, \hat f_{j,i} =-\, q^{a_{j,i}+l}\,\,   \prod_{k\neq j} {g(1)^2\,\,e^{({1\over 2}+{i\pi\over \ln q})\, (1+a_{j,i}-a_{k,i})\ln q}\, \,e^{{(\ln{(q^{1+a_{j,i}-a_{k,i}})})^2\over  2 g_s}}\over g(q^{a_{j,i}-a_{k,i}})\,(1-q^{a_{k,i}-a_{j,i}}) g(q^{a_{k,i}-a_{j,i}})} \,,
\eeqn
where $\hat f_{j,i}$ is independent of the integer $l$. 

The parameters $S_i$ are defined by 
\beqn \label{si}
S_i =  s_{0,i-1}+t_{0,i-1}= s_{j,i-1} -\sum_{k<j} t_{k,i}+\sum_{k\leq j} t_{k,i-1} \,.
\eeqn
The final equality holds for arbitrary $j$ \cite{part_1}.

For $i=0$ and $i=m+1$, we define
\beq
f_0(x) = {1\over \prod_{j=0}^n \prod_{i=1}^d (x-q^{h_i(\alpha_{j,0})})} \,,
\eeq
\beq
f_{m+1}(x) = {1\over \prod_{j=0}^n \prod_{i=1}^d  (x-q^{h_i(\alpha_{j,m+1})})} \,.
\eeq
Notice that if the representations $\vec\alpha_0$ or $\vec\alpha_{m+1}$ are trivial, i.e. $h_i(\alpha_{j,0})=d-i+a_{j,0}$ or $h_i(\alpha_{j,m+1})=d-i+a_{j,m+1}$, we have
\beq
f_0(x) = \prod_{j=0}^n {g(x\,q^{1-a_{j,0}-d})\over x^d\, g(x\,q^{1-a_{j,0}}) }\,,\hspace{1cm} f_{m+1}(x) = \prod_{j=0}^n {g(x\,q^{1-a_{j,m+1}-d})\over x^d\, g(x\,q^{1-a_{j,m+1}}) }
\eeq
respectively.
The functions $f_0$ and $f_{m+1}$ have simple poles
at $x=q^{h_l(\alpha_{j,0})}$ (resp. $x=q^{h_l(\alpha_{j,m+1})}$) for $l=1,\dots,d$, with residue
\beq
\hat f_{j,0;l} = \Res_{q^{h_l(\alpha_{j,0})}} f_0(x) =   {1\over \prod_{j'\neq j} \prod_{i=1}^d (q^{h_l(\alpha_{j,0})}-q^{h_i(\alpha_{j',0})})} \,{1\over  \prod_{i\neq l} (q^{h_l(\alpha_{j,0})}-q^{h_i(\alpha_{j,0})})} \,,
\eeq
\beq
\hat f_{j,m+1;l} = \Res_{q^{h_l(\alpha_{j,m+1})}} f_{m+1}(x) =   {1\over \prod_{j'\neq j} \prod_{i=1}^d (q^{h_l(\alpha_{j,m+1})}-q^{h_i(\alpha_{j',m+1})})} \,{1\over  \prod_{i\neq l} (q^{h_l(\alpha_{j,m+1})}-q^{h_i(\alpha_{j,m+1})})} \,.
\eeq
The $l$ dependence here is more intricate than above, but this will not play any role since the partitions $\alpha_{j,0}$ and $\alpha_{j,m+1}$ are kept fixed, and not summed upon.
  
\bigskip

\begin{figure}[h]
 \centering
 \includegraphics[width=8cm]{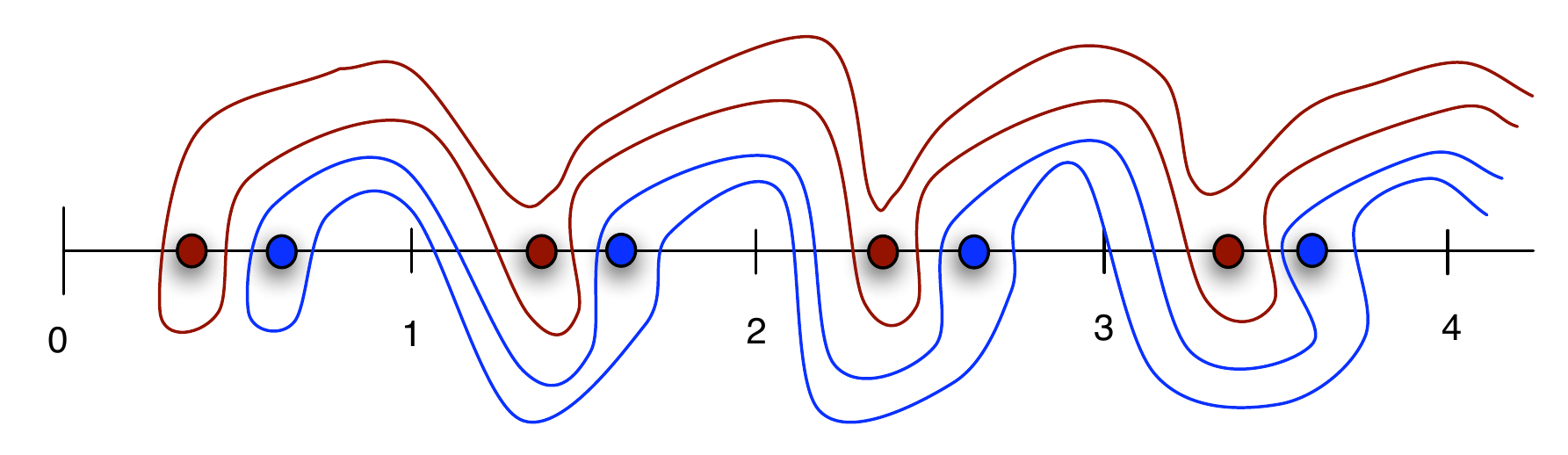}
 \caption{\footnotesize{Two contours surrounding points $a+\mathbb N$ and $b+\mathbb N$, such that  $a-b \notin \mathbb Z$.}}
 \label{contours}
\end{figure}

The integration domains for the matrices $R_i$ are $H_N(\mathbb R_+^N)$, i.e. the set of hermitian matrices  having only positive eigenvalues. For the matrices $M_i, i=1,\dots,m$, the integration domains are $H_N(\Gamma_i)$, where 
\beq
\Gamma_i = \prod_{j=0}^n\, (\gamma_{j,i})^d \,.
\eeq
$\gamma_{j,i}$ is defined as a contour which encloses all points of the form $q^{a_{j,i}+\mathbb N}$, and does not intersect any contours $\gamma_{k,l}$, $(j,i) \neq (k,l)$. For this to be possible, we must require that the differences $a_{j,i}-a_{j',i'}$ be non-integer. The normalized logarithms of two such contours are depicted in figure \ref{contours}.

 We have defined
\beq
H_N(\Gamma_i) = \{ M=U\, \Lambda \, U^\dagger \, , \quad U\in U(N)\, , \,\,\, \Lambda={\rm diag}(\lambda_1,\dots,\lambda_N)\, \in \Gamma_i \} \,,
\eeq
i.e. $H_N(\Gamma_i)$ is the set of normal matrices with eigenvalues on $\Gamma_i$. By definition, the measure on $H_N(\Gamma_i)$ is (see \cite{MehtaBook})
\beqn\label{eqdefdMdUdL}
dM = {1\over N!}\,\, \Delta(\Lambda)^2\,\, dU\,d\Lambda \,,
\eeqn
where $dU$ is the Haar measure on $U(N)$, and $d\Lambda$ is the product of the measures for each eigenvalue along its integration path.

The integration domains for the matrices $M_0$, $M_{m+1}$ are $H_N(\Gamma_0)$, $H_N(\Gamma_{m+1})$ respectively, where 
\beqn  \label{outer_contours}
\Gamma_0 =  (\sum_{j=0}^n \gamma_{j,0})^N \,,
\qquad \quad
\Gamma_{m+1} =  (\sum_{j=0}^n \gamma_{j,m+1})^N \,.
\eeqn

\bibliography{draft_part2}

\providecommand{\href}[2]{#2}\begingroup\raggedright\begin{thebibliography}{10}

\bibitem{part_1}
B.~Eynard, A.-K. Kashani-Poor, and O.~Marchal, ``{A matrix model for the
  topological string I: Deriving the matrix model},''
\href{http://arxiv.org/abs/1003.1737}{{\tt arXiv:1003.1737 [hep-th]}}.

\bibitem{eynloop1mat}
B.~Eynard, ``{Topological expansion for the 1-hermitian matrix model
  correlation functions},''
  \href{http://dx.doi.org/10.1088/1126-6708/2004/11/031}{{\em JHEP} {\bf 11}
  (2004)  031},
\href{http://arxiv.org/abs/hep-th/0407261}{{\tt arXiv:hep-th/0407261}}.

\bibitem{EOFg}
B.~Eynard and N.~Orantin, ``{Invariants of algebraic curves and topological
  expansion},''
\href{http://arxiv.org/abs/math-ph/0702045}{{\tt arXiv:math-ph/0702045}}.

\bibitem{eynhaeq}
B.~Eynard, M.~Marino, and N.~Orantin, ``{Holomorphic anomaly and matrix
  models},'' {\em JHEP} {\bf 06} (2007)  058,
\href{http://arxiv.org/abs/hep-th/0702110}{{\tt arXiv:hep-th/0702110}}.

\bibitem{Orantin}
N.~Orantin, ``{Symplectic invariants, Virasoro constraints and Givental
  decomposition},''
\href{http://arxiv.org/abs/0808.0635}{{\tt arXiv:0808.0635 [math-ph]}}.

\bibitem{CMMV}
L.~Chekhov, A.~Marshakov, A.~Mironov, and D.~Vasiliev, ``{DV and WDVV},''
  \href{http://dx.doi.org/10.1016/S0370-2693(03)00543-4}{{\em Phys. Lett.} {\bf
  B562} (2003)  323--338},
\href{http://arxiv.org/abs/hep-th/0301071}{{\tt arXiv:hep-th/0301071}}.

\bibitem{eynMgnkappa}
B.~Eynard, ``{Recursion between Mumford volumes of moduli spaces},''
  \href{http://arxiv.org/abs/0706.4403}{{\tt arXiv:0706.4403}}.

\bibitem{BKMP}
V.~Bouchard, A.~Klemm, M.~Marino, and S.~Pasquetti, ``{Remodeling the
  B-model},'' \href{http://dx.doi.org/10.1007/s00220-008-0620-4}{{\em Commun.
  Math. Phys.} {\bf 287} (2009)  117--178},
\href{http://arxiv.org/abs/0709.1453}{{\tt arXiv:0709.1453 [hep-th]}}.

\bibitem{marino2}
M.~Marino, ``{Open string amplitudes and large order behavior in topological
  string theory},'' \href{http://dx.doi.org/10.1088/1126-6708/2008/03/060}{{\em
  JHEP} {\bf 03} (2008)  060},
\href{http://arxiv.org/abs/hep-th/0612127}{{\tt arXiv:hep-th/0612127}}.

\bibitem{eynLP}
B.~Eynard, ``{All orders asymptotic expansion of large partitions},''
  \href{http://dx.doi.org/10.1088/1742-5468/2008/07/P07023}{{\em J. Stat.
  Mech.} {\bf 0807} (2008)  P07023},
\href{http://arxiv.org/abs/0804.0381}{{\tt arXiv:0804.0381 [math-ph]}}.

\bibitem{MarshakovNekrasov}
A.~Marshakov and N.~Nekrasov, ``{Extended Seiberg-Witten theory and integrable
  hierarchy},'' {\em JHEP} {\bf 01} (2007)  104,
\href{http://arxiv.org/abs/hep-th/0612019}{{\tt arXiv:hep-th/0612019}}.

\bibitem{KlemmSulkowski}
A.~Klemm and P.~Sulkowski, ``{Seiberg-Witten theory and matrix models},''
  \href{http://dx.doi.org/10.1016/j.nuclphysb.2009.04.004}{{\em Nucl. Phys.}
  {\bf B819} (2009)  400--430},
\href{http://arxiv.org/abs/0810.4944}{{\tt arXiv:0810.4944 [hep-th]}}.

\bibitem{Sulkowski}
P.~Sulkowski, ``{Matrix models for 2* theories},''
\href{http://arxiv.org/abs/0904.3064}{{\tt arXiv:0904.3064 [hep-th]}}.

\bibitem{Zhou}
J.~Zhou, ``{Local Mirror Symmetry for One-Legged Topological Vertex},''  (2009)
   , \href{http://arxiv.org/abs/0910.4320}{{\tt arXiv:0910.4320 [math.AG]}}.

\bibitem{Lin}
L.~Chen, ``{Bouchard-Klemm-Marino-Pasquetti Conjecture for $\mathbb{C}^3$},''
  (2009)  , \href{http://arxiv.org/abs/0910.3739}{{\tt arXiv:0910.3739
  [math.AG]}}.

\bibitem{BouchardMarino}
V.~Bouchard and M.~Marino, ``{Hurwitz numbers, matrix models and enumerative
  geometry},''
\href{http://arxiv.org/abs/0709.1458}{{\tt arXiv:0709.1458 [math.AG]}}.

\bibitem{BEMS}
G.~Borot, B.~Eynard, M.~Mulase, and B.~Safnuk, ``{A matrix model for simple
  Hurwitz numbers, and topological recursion},''
\href{http://arxiv.org/abs/0906.1206}{{\tt arXiv:0906.1206 [math-ph]}}.

\bibitem{EMS}
B.~Eynard, M.~Mulase, and B.~Safnuk, ``{The Laplace transform of the
  cut-and-join equation and the Bouchard-Marino conjecture on Hurwitz
  numbers},''
\href{http://arxiv.org/abs/0907.5224}{{\tt arXiv:0907.5224 [math.AG]}}.

\bibitem{OSY}
H.~Ooguri, P.~Sulkowski, and M.~Yamazaki, ``{Wall Crossing As Seen By Matrix
  Models},''
\href{http://arxiv.org/abs/1005.1293}{{\tt arXiv:1005.1293 [hep-th]}}.

\bibitem{HoriVafa}
K.~Hori and C.~Vafa, ``{Mirror symmetry},''
\href{http://arxiv.org/abs/hep-th/0002222}{{\tt arXiv:hep-th/0002222}}.

\bibitem{AKV}
M.~Aganagic, A.~Klemm, and C.~Vafa, ``{Disk instantons, mirror symmetry and the
  duality web},'' {\em Z. Naturforsch.} {\bf A57} (2002)  1--28,
\href{http://arxiv.org/abs/hep-th/0105045}{{\tt arXiv:hep-th/0105045}}.

\bibitem{AKMV}
M.~Aganagic, A.~Klemm, M.~Marino, and C.~Vafa, ``{The topological vertex},''
  \href{http://dx.doi.org/10.1007/s00220-004-1162-z}{{\em Commun. Math. Phys.}
  {\bf 254} (2005)  425--478},
\href{http://arxiv.org/abs/hep-th/0305132}{{\tt arXiv:hep-th/0305132}}.

\bibitem{IqbalKashaniPoor}
A.~Iqbal and A.-K. Kashani-Poor, ``{The vertex on a strip},'' {\em Adv. Theor.
  Math. Phys.} {\bf 10} (2006)  317--343,
\href{http://arxiv.org/abs/hep-th/0410174}{{\tt arXiv:hep-th/0410174}}.

\bibitem{MehtaBook}
M.~L. Mehta, {\em Random matrices}, vol.~142 of {\em Pure and Applied
  Mathematics (Amsterdam)}.
\newblock Elsevier/Academic Press, Amsterdam, third~ed., 2004.

\bibitem{DFGZJ}
P.~Di~Francesco, P.~H. Ginsparg, and J.~Zinn-Justin, ``{2-D Gravity and random
  matrices},'' \href{http://dx.doi.org/10.1016/0370-1573(94)00084-G}{{\em Phys.
  Rept.} {\bf 254} (1995)  1--133},
\href{http://arxiv.org/abs/hep-th/9306153}{{\tt arXiv:hep-th/9306153}}.

\bibitem{Eynchain}
B.~Eynard, ``{Master loop equations, free energy and correlations for the chain
  of matrices},'' {\em JHEP} {\bf 11} (2003)  018,
\href{http://arxiv.org/abs/hep-th/0309036}{{\tt arXiv:hep-th/0309036}}.

\bibitem{EPrats}
B.~Eynard and A.~P. Ferrer, ``{Topological expansion of the chain of
  matrices},''
\href{http://dx.doi.org/10.1088/1126-6708/2009/07/096}{{\em JHEP} {\bf 07}
  (2009)  096}.

\bibitem{toappear}
B.~Eynard and O.~Marchal. {work in progress}.

\bibitem{MarcoPaths}
M.~Bertola, ``Biorthogonal polynomials for 2-matrix models with semiclassical
  potentials,'' {\em J.APPROX.THEORY 144} {\bf 2} (2007)  162.

\bibitem{EOreview}
B.~Eynard and N.~Orantin, ``{Algebraic methods in random matrices and
  enumerative geometry},''
\href{http://arxiv.org/abs/0811.3531}{{\tt arXiv:0811.3531 [math-ph]}}.

\bibitem{BDE}
G.~Bonnet, F.~David, and B.~Eynard, ``{Breakdown of universality in multi-cut
  matrix models},'' \href{http://dx.doi.org/10.1088/0305-4470/33/38/307}{{\em
  J. Phys.} {\bf A33} (2000)  6739--6768},
\href{http://arxiv.org/abs/cond-mat/0003324}{{\tt arXiv:cond-mat/0003324}}.

\bibitem{Eynhardedge}
B.~Eynard, ``{Loop equations for the semiclassical 2-matrix model with hard
  edges},'' {\em J. Stat. Mech.} {\bf 0510} (2005)  P006,
\href{http://arxiv.org/abs/math-ph/0504002}{{\tt arXiv:math-ph/0504002}}.

\bibitem{ChekhovHardedges}
L.~Chekhov, ``{Matrix models with hard walls: Geometry and solutions},'' {\em
  J.Phys.A} {\bf 39} (2006)  8857--8894,
  \href{http://arxiv.org/abs/0602013}{{\tt arXiv:0602013 [hep-th]}}.

\bibitem{Fay}
J.~D. Fay, {\em Theta functions on {R}iemann surfaces}.
\newblock Lecture Notes in Mathematics, Vol. 352. Springer-Verlag, Berlin,
  1973.

\bibitem{KK}
A.~Kokotov and D.~Korotkin, ``Tau-function on Hurwitz spaces,'' {\em
  Mathematical Physics, Analysis and Geometry} {\bf 7} (2004)  47--96.

\bibitem{EKK}
B.~Eynard, A.~Kokotov, and D.~Korotkin, ``{Genus one contribution to free
  energy in hermitian two- matrix model},''
  \href{http://dx.doi.org/10.1016/j.nuclphysb.2004.06.031}{{\em Nucl. Phys.}
  {\bf B694} (2004)  443--472},
\href{http://arxiv.org/abs/hep-th/0403072}{{\tt arXiv:hep-th/0403072}}.

\bibitem{Johansson}
K.~Johansson, ``The arctic circle boundary and the {A}iry process,''
  \href{http://dx.doi.org/10.1214/009117904000000937}{{\em Ann. Probab.} {\bf
  33} (2005) no.~1, 1--30}.

\bibitem{Jost}
J.~Jost, {\em Compact {R}iemann surfaces}.
\newblock Springer-Verlag, Berlin, second~ed., 2002.
\newblock An introduction to contemporary mathematics.

\bibitem{Witten_Phases}
E.~Witten, ``{Phases of N = 2 theories in two dimensions},''
  \href{http://dx.doi.org/10.1016/0550-3213(93)90033-L}{{\em Nucl. Phys.} {\bf
  B403} (1993)  159--222},
\href{http://arxiv.org/abs/hep-th/9301042}{{\tt arXiv:hep-th/9301042}}.

\bibitem{KonishiMinabe}
Y.~Konishi and S.~Minabe, ``{Flop invariance of the topological vertex},''
  \href{http://dx.doi.org/10.1142/S0129167X08004546}{{\em Int. J. Math.} {\bf
  19} (2008)  27--45},
\href{http://arxiv.org/abs/math/0601352}{{\tt arXiv:math/0601352}}.

\bibitem{Cavalieri}
R.~Cavalieri. {private communications}.

\bibitem{AIM}
{\em {Recursion structures in topological string theory and enumerative
  geometry}}.
\newblock June 8 to June 12, 2009, at the American Institute of Mathematics
  (AIM), Palo Alto, California.

\end{thebibliography}\endgroup
\bibliographystyle{utcaps}

\end{document}